\documentclass[12pt]{elsarticle}
\graphicspath{{Figures/}}
\usepackage{amsmath,amssymb,amsthm,amsfonts}
\usepackage{caption}
\usepackage[title]{appendix}
\usepackage{xcolor}
\setcitestyle{authoryear,open={(},close={)}}

\usepackage{graphicx,epsfig}
\usepackage{psfrag}
\usepackage{xcolor}
\usepackage[hmargin=1in,vmargin=1in]{geometry}
\usepackage{stabular,morefloats}
\usepackage{textcomp}
\usepackage{titlesec}
\usepackage{colortbl}
\usepackage{xcolor}
\usepackage{multirow}
\usepackage{mathrsfs}
\usepackage{tabularx}
\usepackage{comment}
\usepackage{hyperref}
\usepackage{multirow,tabularx,afterpage,fancyhdr}
\usepackage{placeins}
\titleformat{\subsubsection}
  {\normalfont\selectfont}{\thesubsubsection}{1em}{}
\numberwithin{equation}{section}
\usepackage{float}
\usepackage{xcolor}
\usepackage{soul}
\definecolor{lightblue}{rgb}{0,0,0.1}
\sethlcolor{lightblue}

\usepackage[subrefformat=parens,labelformat=parens]{subfig}

\usepackage{subcaption}

\colorlet{Changes@Color}{black}
\makeatletter
\renewcommand{\@biblabel}[1]{[#1]\hfill}
\makeatother

\makeatletter
\def\ps@pprintTitle{%
 \let\@oddhead\@empty
 \let\@evenhead\@empty
 \def\@oddfoot{}%
 \let\@evenfoot\@oddfoot}
\makeatother

\begin{document}

\title{On modeling fracture of soft polymers}
\author[main]{Aditya Konale}
\author[main,main2]{Vikas Srivastava\corref{cor1}}
\ead{vikas\_srivastava@brown.edu}
\cortext[cor1]{Corresponding author}
\address[main]{School of Engineering, Brown University, Providence, RI 02912, USA}
\address[main2]{Institute for Biology, Engineering and Medicine, Brown University, Providence, RI 02912, USA}

\begin{abstract}Soft polymers are ubiquitous materials in nature and as engineering materials with properties varying from rate-independent to rate-dependent. Current fracture toughness measures are non-unique for rate-dependent soft materials for varying loading profiles and specimen geometries. Works on modeling fracture in rate-dependent soft polymers are limited to specific pre-cracked geometries. There is no generally agreed-upon model for the fracture of soft polymers. We propose and show that a critical value of stress work can be used as a measure of fracture resistance for a certain class of soft polymers. We develop a damage model to predict fracture in soft polymers. In the model, the energetic part of the critical stress work is proposed as a damage initiation criterion that has the ability to capture failure surfaces. The damage growth is modeled through a generalized gradient-damage framework. The fracture model is validated for both elastomers and viscous soft polymers by comparing model predictions against experimental results for different materials (ethylene propylene diene monomer - EPDM, EPS25 vitrimer, styrene butadiene rubber - SBR, natural rubber - NR, and polyborosiloxane - PBS), a variety of specimen geometries, and loading conditions. The model can predict key physical phenomena such as brittle and ductile responses and different fracture profiles. The microstructural quantities, such as subchain dissociation energy during the fracture of polymers, can be predicted from the macroscopic model parameters.
\end{abstract}


\maketitle
\newpage 
\section{Introduction}\label{sec1}

Soft polymers are defined as a broad class of polymers with chain networks formed through permanent or dynamic chemical or mechanical crosslinks, leading to elastic modulus in the range of a few kilopascals to tens of megapascals. This class encompasses elastomers, viscoelastic and elastic-viscoplastic polymers, polymer gels,\footnote{The polymer gels here are restricted to those where loading timescales are significantly faster than the solvent diffusion timescale.} and biological tissues. Their physical forms can vary from very viscous, slowly flowing soft solids \citep{KonalePBS} to elastomeric rubber-like materials \citep{Hyperelastic2}.

Such polymers form a key class of engineering materials with a broad range of applications varying from biomedical drug delivery \citep{shukla2020, XZhaodrugdelivery, ALKHOURY2024113086, LIdrugdelivery} and health monitoring \citep{Softpolymerapp1, Softpolymerapp2, Hwangpollenelectronics} to impact protection \citep{Zongetal(2021)rheometryref, Tangetal(2022)B:Ocrosslinkrubberyref}, actuation \citep{Yuk2017, PBSsensingactuators, Yingliactuator}, and adhesives \citep{Yuk2016, XZhaoadhesive}. These polymers possess significant variations in mechanical properties due to their underlying crosslinking mechanisms. Permanent crosslinks result in rate-independent behavior with high elasticity like elastomers \citep{Hyperelastic1, Hyperelastic2, UPADHYAY2021103949}. On the other hand, dynamic, reversible crosslinks lead to significantly rate-dependent mechanical behavior \citep{Dynamiccrlnk1, Kurkinetal(Synthesis)}. In the case of dynamically crosslinked polymers, longer loading timescales lead to highly viscous fluid-like behavior, while shorter loading timescales result in elastomer-like behavior \citep{KonalePBS, Dynamiccrlnk2}. The dynamic crosslinks can provide autonomous self-healing of small fractures \citep{Dynamiccrlnk2, Yingli1}. 
Hence, depending on the nature of crosslinks in a soft polymer, its behavior can vary between the extremes of purely elastic to highly viscous response. Current fracture toughness measures based on energy release rate or crack tip stress intensity are non-unique when evaluated for rate-dependent soft materials. They vary with the loading profile and specimen geometry under consideration \citep{Songfracture, Fracturechallenges1, Fracturechallenges2, Npm2units1}. It will be of value to have a general measure in a continuum sense that is not geometry and rate-dependent.

Significant experimental studies have focused on the fracture of elastomers and polymer gels \citep{Lakethomas1967, Lake1995elastomers, CHENflawsensitivity, Cretonperspective, Fracturechallenges2, Kimhydrogelfatiguefracture, TANG201724, SHRIVASTAVA2023104524}. Recent modeling efforts, which are limited only to elastomers based on the gradient-damage framework, can be broadly separated into three approaches. First, considers a variational approach by minimizing a non-linear energy function to predict crack nucleation without explicit criteria \citep{Variationalpf1, Variationalpf2}. The drawback of the variational approach, being unable to capture experimentally observed failure surfaces, has been noted in \citep{LOPEZPAMIES2025117520}. The second approach for elastomers directly considers the underlying Euler-Lagrange equations and uses independent stress-based criteria for damage nucleation from the bulk \citep{KUMAR1, KUMAR2, LOPEZPAMIES2025117520}. It uses the Griffith energy competition as a descriptor of nucleation from a large pre-existing crack. The third approach is based on the method of virtual power. In this approach, the undamaged free energy density function, which is the driving force for the initiation and evolution of damage, is formulated in two ways. One considers it to be entirely entropic following the ideal chain assumption \citep{MIEHE1, Miehe2, Holzapfelarterial, WU2016596}. The other incorporates an energetic contribution to the free energy density function due to backbone bond or crosslink stretching (whichever is weaker) along with the conventional entropic part \citep{Maochainscission, TALAMINI, Maocrosslinkfailure, Maogels, LI2020193}. This is motivated by the scaling law for elastomers proposed by Lake and Thomas \citep{LakeandThomas} relating the Griffith energy release rate to the binding energy between monomer units (or of crosslinks \citep{Maocrosslinkfailure}) among other microscopic parameters. However, all these models are limited to elastomers only, while the focus of the paper is to develop a general, unified damage and fracture model for rate-dependent, viscous polymers that is also applicable to elastomers.

The rate-dependent fracture of polymers has been of significant interest, and various relevant experimental studies have been published \citep{ROSAKIS2000, ArunShukla2010, MacDonald2020, LEBLANC2020, LeBlanc2018, SrivastavaSHPB1, SrivastavaSHPB2, Gsubhash1, Gsubhash2, ZHENG2022, malhotra2021, Brown2002, MALITO2019, Burla2020, Liu2023, COX2005565, Guzekpolymerinterfacefracture, XuMom4}. The rate-dependent fracture of polymers that contain dynamic crosslinks has been recently experimentally studied \citep{Polyampholytefracturestudy, ZehnderkineticfracturePVA, Songfracture, ZHAOfractureDNgels2, Chen2023ratedependentfracture, Baiselfrecoverygelfracture, LiuMom}. Mayumi and co-workers \citep{Mayumifracture} conducted notched tension tests on Polyvinyl alcohol gels and noticed a significant decrease in failure stretch with increasing loading rates. Shen and co-workers \citep{Shenmodeling} performed notched pure shear tests on EPS25 vitrimer with multi-step loading profiles. The authors observed loading rate-dependent flow-fracture (crack blunting-advancing) transitions. Qi and co-workers \citep{QIdisspmappingPA} presented a method to map the dissipation field during crack propagation in viscous soft polymers and enable individual evaluation of the intrinsic and dissipative components of fracture toughness. The authors evaluated the stress and, consequently, the total energy density provided by mechanical work at each material point using a non-linear constitutive model and deformation field measurements. The dissipation field was then obtained by subtracting the free energy density field evaluated using the constitutive model from the total energy density field. 

The modeling of the rate-dependent fracture in dynamically crosslinked soft polymers and viscous polymers is limited and has been restricted to crack tip stress field analyses \citep{HuiJMPSfracture, Zehndertipstressfield, HuiModeIII, ZehnderkineticfracturePVA} or crack tip driving force-based approach \citep{Shenmodeling2, Shenmodeling, CIAVARELLA2024104564} for specific pre-cracked geometries. The crack tip stress field asymptotic analyses use the deformation and stress fields near the crack tip to predict fracture. The crack tip driving force approach relates the crack propagation velocity to the energy release rate. Models with the ability to predict crack nucleation, propagation, and coalescence in arbitrary, complex geometries are lacking and are necessary for solving a general class of problems. The gradient-damage approach used for modeling mechanical damage and complete failure in a broad range of materials is a suitable choice for this purpose \citep{Miehefirstpaper, Anandquasibrittle, Anand(2021)flowrule, KUHL2000, Shabnamphasefield, Guocytoskeletalpf, Dolbowpf2, BeheraMom2, DuMom3}. There have been informative discussions in the literature on the limitations of certain damage and fracture computation methods and models \citep{LOPEZPAMIES2025117520, Bazant1, Bazant3}. Sometimes, the limitations of constitutive damage models and the limitations of computational methods are not distinguished separately. For the assessment of overall numerical predictive abilities, the constitutive and computational strengths and weaknesses should be separately evaluated.  

Critical gaps in modeling and predicting damage initiation, growth, and fracture of soft polymers are (i) the establishment of a unique physical parameter to quantify fracture resistance of soft polymers and (ii) the development of a unified model to predict damage initiation and growth which is generally applicable for soft polymers that range from viscous polymers to elastomers and is independent of loading history and specimen geometry. 

For rate-dependent soft polymers, we considered Polyborosiloxane (PBS). PBS is a dynamically crosslinked, rate-stiffening soft polymer that is highly viscous at long timescales with a significantly rate-dependent mechanical response \citep{Lietal(2014)B:Ocrosslinkref, Tangetal(Synthesis), Zhaoetal(2018)IHE, ZHAOSSGReview, Kim2022SSG}. PBS flows under its weight over long timescales (minutes) and exhibits a rubber-like response with a significant shape recovery upon deformation when subjected to high loading rates. The rate-dependent fracture response of PBS has not been experimentally studied and corresponding models have not been developed, which are critical given viscous polymers such as PBS's broad applications, which include self-healing sensing and actuation \citep{PBSsensingactuators, Wangetal(2018)AnovelmagneticSSGwithselfhealingability, Liuetal(2019)B:Ocrosslinkref} and flexible impact mitigation \citep{TU2023116811, Myronidisetal(2022)rheometryref}.

Below are the main contributions of this paper: 
\begin{enumerate}[(i)]
\item We have proposed and shown that a critical value of stress work $\mathit{W}_{cr}$ can be an approximately unique measure of fracture resistance for a certain class of soft polymers.\footnote{$W_{cr}$ is a general definition in the continuum sense. The work of fracture can be considered as a special case of $W_{cr}$. For a uniaxial tensile loading of an unnotched specimen, $\mathit{W}_{cr}$ is equivalent to the standard modulus of toughness \citep{modoftoughness1, modoftoughness2} or work of fracture [defined as the area under the engineering stress-strain curve in an unnotched uniaxial tension test till failure] \citep{CHENflawsensitivity, Worktorupture2, Fracturepzonesizesensitive2}.} It quantifies the total energy dissipation per unit volume for the complete failure of an arbitrary material point under a loading mode when it is subjected to loading rates sufficient to cause damage initiation. Complete failure refers to the inability to support any mechanical stress. $\mathit{W}_{cr}$ can be numerically evaluated for general deformations, of which the homogeneous deformations are special cases, where it can be directly obtained from the experiments.
\item We have developed a model to predict damage initiation, damage growth, and complete failure in soft polymers using a multi-mechanism generalized gradient-damage framework based on the method of virtual power, where the material strength is accounted for.\footnote{This method uses explicit criteria for damage initiation in contrast to the variational phase field method, which minimizes an energy function to predict damage initiation. The limitations of the variational phase field method have recently been discussed in \citep{LOPEZPAMIES2025117520}.} The free energy density contribution from network resistance is formulated considering crosslink relaxation and formation. As motivated by our previous work \citep{KonalePBS}, the number of chain segments that contribute to the free energy and their individual stretches vary with time, as governed by the underlying crosslink kinetics.\footnote{This form of free energy reflects microstructural physics and is thermodynamically consistent. The free energy density functions with direct rate-dependence proposed in some of the literature are thermodynamically inconsistent.} The damage initiation criterion $\psi^{+}_{cr}$ (\emph{energetic contribution to $\mathit{W}_{cr}$}) in the model can be evaluated using a one-dimensional numerical implementation of deformation-only constitutive model and homogeneous deformation experiment (e.g., simple tension test) results.
\item We validated the predictive capabilities of the fracture model for three elastomers (EPDM, SBR, NR), a viscous soft polymer (EPS25 vitrimer), and PBS.
    
\end{enumerate}

\section{Fracture in soft polymers}

\begin{figure}[hbt!]
    \begin{center}
		\includegraphics[width=\textwidth]{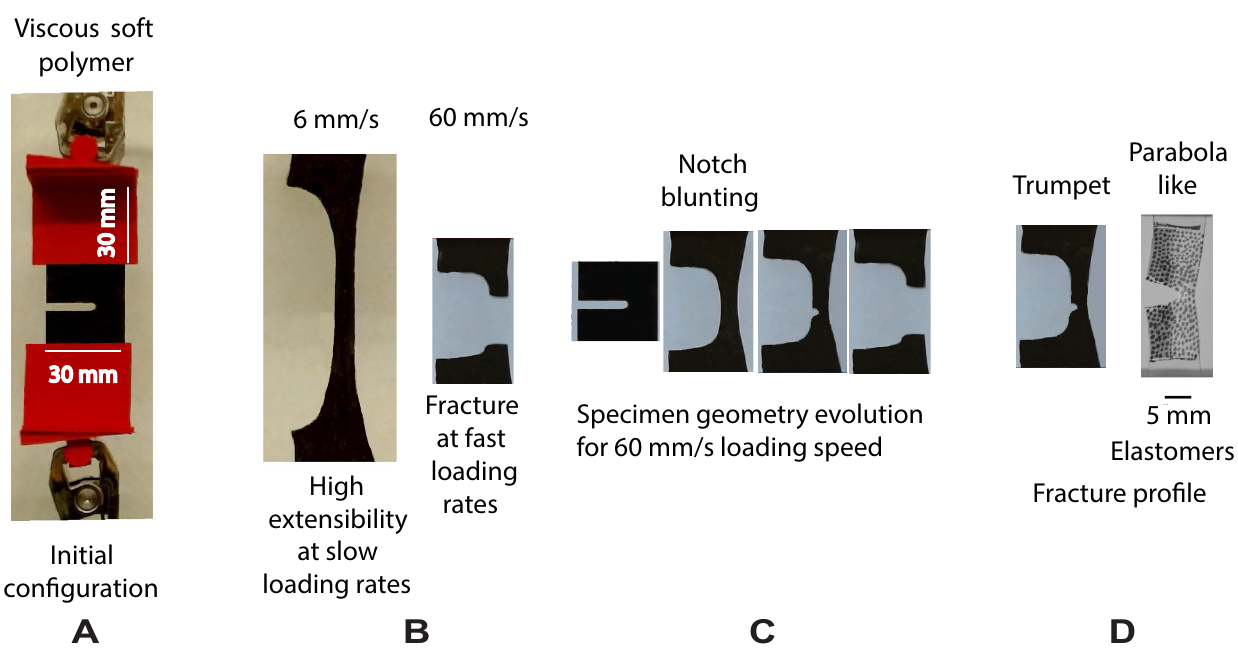}
	\end{center}
    \caption{\textbf{Fracture phenomenon in soft polymers.} {(\textbf{A}) The geometry of a notched PBS specimen and the experiment schematic. The thickness of the specimen is 3.2 mm. PBS was colored for visualization. (\textbf{B}) The high extensibility of the notched PBS specimen when stretched at a slow loading speed (6 mm/s). The fracture of the specimen at a fast loading speed (60 mm/s). (\textbf{C}) The specimen geometry evolution till complete failure for the 60 mm/s loading speed. Significant notch blunting occurs before notch propagation. (\textbf{D}) The trumpet fracture profile (formation of a second local notch ahead of the original blunted notch) in the rapidly stretched notched PBS specimen. A single parabola-like fracture profile experimentally observed in PDMS elastomer (image from J. Lee and co-workers (2024) \citep{Parabolicshape2} reproduced with Creative Commons [CC BY-NC-ND 4.0 - https://creativecommons.org/licenses/by-nc-nd/4.0/] copyright license).   
    }} 
    \label{Motivation}
\end{figure}

Notches and cracks in soft polymers undergo a significant amount of blunting before fracture initiation \citep{MIEHE1, Polyampholytefracturestudy}. Hence, geometric non-linearity due to the large deformation corresponding to blunting has to be accounted for along with material non-linearity. Figure \ref{Motivation} depicts key observations during the deformation and fracture of PBS. To fabricate PBS, hydroxy-terminated Polydimethylsiloxane (PDMS, Sigma Aldrich 481963, 750 cSt) and Boric Acid (BA, Sigma Aldrich B0394, ACS reagent, $\ge$ 99.5$\%$) were mixed in the ratio 160:1 by weight. The mixture was stirred and then heated at 120$^{\text{o}}$C for 72 hrs to obtain PBS \citep{KonalePBS}. The mass density of PBS was measured to be 1100 kg m$^{-3}$. A notched PBS specimen in Figure \ref{Motivation}A was stretched at slow (6 mm/s) and fast (60 mm/s) rates.  All tension tests on PBS in this work were conducted using the UniVert Mechanical Test System (CellScale) and at 23$^{\text{o}}$C. Three replicates were performed for each experimental result reported. We applied 3D-printed Polylactic acid (PLA) custom fixtures to avoid creeping of PBS under the clamping pressure of conventional spring-loaded grips. In the custom fixtures shown in Figure \ref{Motivation}A, parts of the specimen outside the gauge length are glued to the fixture's interior faces using a cyanoacrylate adhesive. PBS's rate-dependent fracture response is highlighted in Figure \ref{Motivation}B through a transition from high extensibility at a slow deformation rate without fracture to fracture with low extensibility at a fast loading rate. Figure \ref{Motivation}C shows the specimen geometry evolution with stretching load through damage evolution and complete failure at the 60 mm/s loading rate. The high extent of notch blunting before notch propagation can be observed in Figure \ref{Motivation}C. The presence of viscous effects in soft polymers results in a trumpet-like early fracture profile (formation of second local notch ahead of the original blunted notch) \citep{Viscoelastictrumpet1, Viscoelastictrumpet2}. Figure \ref{Motivation}D shows deformation profiles prior to complete fracture in notched PBS and notched Polydimethylsiloxane (PDMS) elastomer \citep{Parabolicshape2} specimens. Compared to the trumpet profile for PBS, the elastomer exhibits a single parabola-like fracture profile. The single parabola-like fracture profile for elastomers has been well documented in literature \citep{Parabolaelastic1, Parabolaelastic2}. A fracture model for soft polymers such as the one proposed in this work, in addition to accurately predicting the failure loads, should predict these significantly varying physical phenomena observed in experiments.

\section{A fracture resistance measure for soft polymers}
\label{Critical stress work discussion}

In permanently crosslinked soft polymers that fracture by crosslink failure, i.e., polymer chain backbone bonds are stronger than crosslinks, it can be assumed that all crosslinks at a material point fail simultaneously \citep{Maocrosslinkfailure}. The weak crosslinks in dynamically crosslinked soft polymers are constantly breaking and reforming, even in an undeformed state. This dissociation rate is amplified by stretching of the corresponding subchains with respect to the stress-free subchain formation configuration \citep{Songfracture, HuiPA1}. Subchains denote parts of chain molecules between two crosslinks connected to the elastically active network. All subchains in a volume element in a dynamically crosslinked soft polymer can be broken simultaneously only when they are stretched equally to the level corresponding to stretch-induced dissociation. This scenario, however, can occur only at infinitely fast rates, which is not practically feasible. At adequate fast rates, a sufficient number of subchains surviving from close to time $t=0$ will dissociate from the elastically active network, creating voids. These voids will coalesce and lead to the complete failure of a volume element \citep{Shenmodeling}. A measure of fracture resistance here can thus be thought of as a critical value of input energy density beyond which network stretching and subsequent damage in the form of stress-accelerated subchain dissociation becomes more dominant than network reconfiguration. 

Stress work is defined as 

\begin{equation}
    \mathit{W}({X}_{i}, t) = \int_{0}^{t} \frac{1}{2} \: {S}_{ij} \dot{{C}}_{ij} dt,
\label{Stress work definition}
\end{equation}

\noindent quantifies the total energy density per unit referential volume supplied to a material point ${X}_{i}$ due to the work done by external macroscopic forces in the scenarios where kinetic energy is negligible. $S_{ij}$, $C_{ij}=F_{ki} F_{kj}$ are the second Piola-Kirchoff stress and the right Cauchy-Green deformation tensors, respectively, with $F_{ij}$ being the deformation gradient tensor. $\mathit{W}_{cr}$ is defined as the value of $\mathit{W}$, i.e., the total energy per unit referential volume that needs to be supplied for the complete failure of an arbitrary material point subjected to sufficiently fast loading rates for damage initiation. Irrespective of the strain rate value under consideration, the same number of bonds have to be broken for the complete failure of a material point. A common loading mode-based response asymmetry observed in soft polymers is tension-compression asymmetry \citep{TCA1, TCA2}. $\mathit{W}_{cr}$'s value can then be different in tension and compression for these soft polymers. However, for a given loading mode (i.e., tension, compression, etc.), the hypothesis is that $\mathit{W}_{cr}$ is a unique value with respect to loading rates and specimen geometries for a certain class of soft polymers.\footnote{Soft polymers, including elastomers, can exhibit near incompressibility \citep{Bulkmoduluselastomers1, Bulkmoduluselastomers2, Bulkmoduluselastomers3}. For the case of pure hydrostatic tension, as incompressibility is approached, the deformation until fracture will be small. But $W_{cr}$ will not be zero under this loading as the ground-state bulk modulus ($K$) of a nearly incompressible polymer is quite large, resulting in finite $W_{cr}$. For the volumetric part of free energy density ($\psi$), if the simple functional form of $\frac{1}{2}K\ {(J-1)}^2$ is considered, then under purely dilatational deformation, $\psi=\frac{1}{2}K\ {(J-1)}^2$, with $J$=det$F_{ij}$. For a purely elastic polymer, $W=\psi$. As an example, if the ground state Young’s modulus and Poisson's ratio of such a nearly incompressible polymer are taken as 2 MPa and 0.4999, respectively, $K$ will be 3.3 GPa. Further, if $J_{cr}$ under purely tensile dilatational deformation is arbitrarily considered to have a very low value of 1.001, then we get $W_{cr}$=1.6 kPa, \emph{which is a significant non-zero value.} This can also be attributed to the non-zero energy density needed for the dissociation of a finite number of bonds, leading to the failure of a material point.} This hypothesis's motivation comes from the earlier argument, along with the following presented experimental observations. It can be understood mathematically on a high level by considering a material represented by a Maxwell element. At slow strain rates, the spring (representing the load-bearing bonds) is not stretched significantly as the dashpot sustains most of the deformation. Consider the reasonable assumption of bond energies being independent of strain rate. Then, with increasing strain rate, the springs (bonds) start stretching significantly, and there will be a specific point (quantified by $W_{cr}$) at which complete failure of the spring (bonds) occurs. Although these arguments may not hold for the Kelvin-Voigt (KV) element, it can be noted that the KV element is not a good representation of most materials. However, the above arguments can be easily extended to the more applicable rheological material idealizations of Standard Linear Solid (SLS) or multiple Maxwell elements in parallel (generalized Maxwell). This can be attributed to the Maxwell elements (or the spring in SLS) in parallel, deforming independently. Hence, each element and the corresponding mechanism of deformation will have its own unique critical energy density value quantifying its damage. $W_{cr}$ can then be interpreted as the sum of these critical energy density values for all mechanisms.

$\mathit{W}_{cr}$ can be evaluated using relatively simple experiments involving homogeneous deformation and without constitutive models. The unit of $\mathit{W}_{cr}$ is J/m$^3$ in contrast to J/m$^2$ for the conventional fracture toughness measures \citep{Npm2units1, Npm2units2}. The standard energy release rate \citep{ERRdef1, ERRdef2} and the essential work of fracture \citep{Npm2units2} are the measures of loss of energy per unit area of crack growth. They usually depend on the constraints created at the crack tip by the amount of material, i.e., specimen dimensions considered \citep{ERRthdepnd1, ERRthdepnd2, Npm2units1}. On the contrary, $\mathit{W}_{cr}$ considers the energy density associated with the complete failure of an arbitrary material point. Hence, the value of $\mathit{W}_{cr}$ should be independent of specimen dimensions. For very soft materials, feasible experiments often involve inhomogeneous deformations \citep{RIFT, Puncturefracture}. $W_{cr}$'s general definition allows its evaluation in arbitrary specimen geometries, contrary to conventional fracture toughness measures, which require specific geometries.

\begin{figure}[hbt!]
    \begin{center}
		\includegraphics[width=0.74\textwidth]{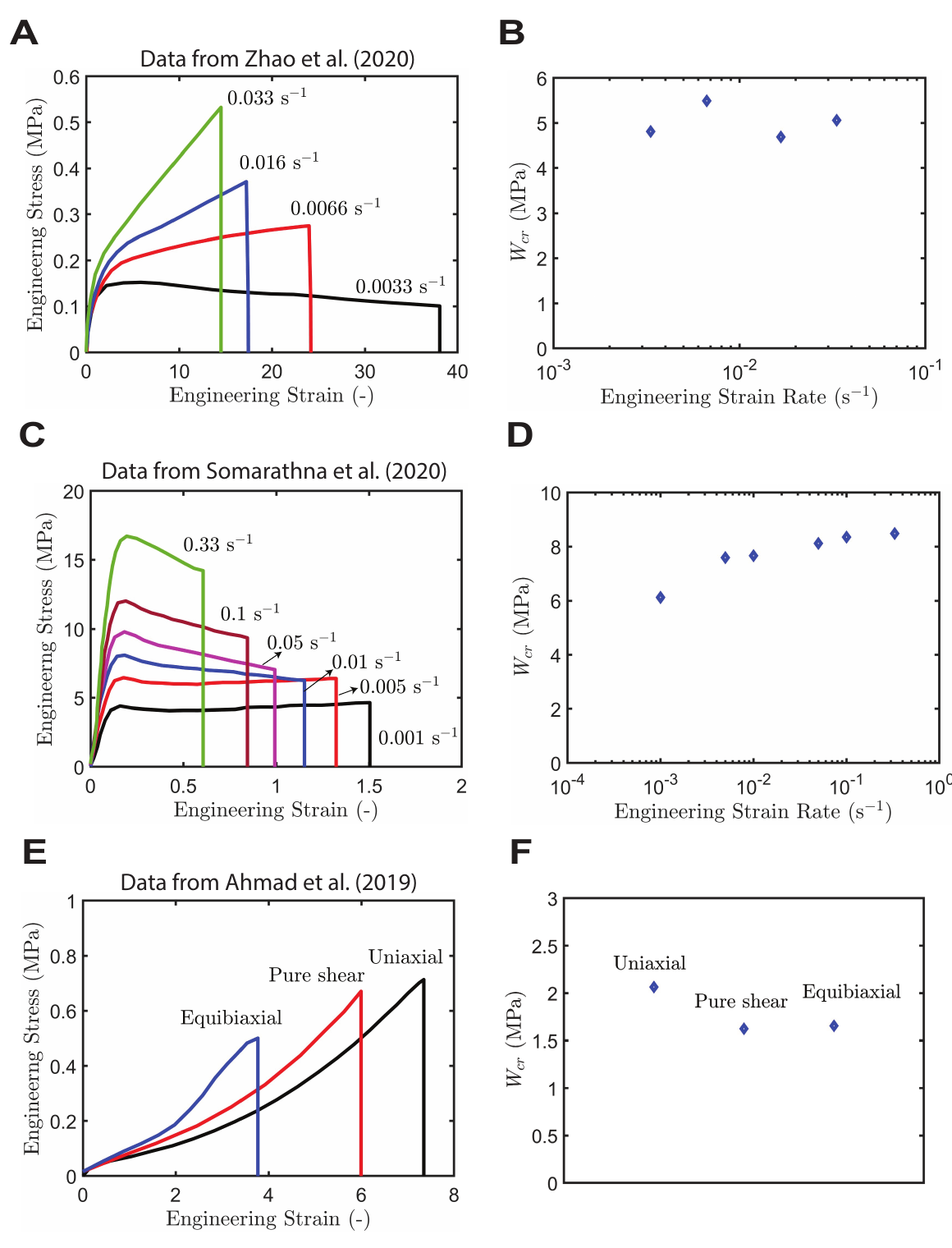}
	\end{center}
    \caption{{\textbf{Approximately constant $W_{cr}$ for broad loading rates and loading modes}. (\textbf{A}) The engineering stress-strain curves at different strain rates for a rate-dependent viscous soft polymer with dynamic crosslinks (TFPM-PDMS-25000) from \citep{fastdamagevolution1}. (\textbf{B}) $\mathit{W}_{cr}$ evaluated as the area under the stress-strain curve is approximately constant over the decade of strain rate considered. (\textbf{C}), (\textbf{D}) The engineering stress-strain curves for polyurethane from \citep{Wcrexample2} and roughly constant $\mathit{W}_{cr}$ evaluated over two decades of strain rates, respectively. (\textbf{E}) The Uniaxial tension, Pure shear, and Equibiaxial tension engineering stress-strain curves for Ecoflex elastomer at a stretch rate of 0.3 s$^{-1}$ from \citep{AHMADdiffloadmodes}. (\textbf{F}) Ecoflex's approximately constant $\mathit{W}_{cr}$ evaluated for three different loading modes considered.}}
    \label{Wcrexamples}
\end{figure}

We now utilize experimental stress-strain curves for a few soft polymers to demonstrate the negligible dependence of $\mathit{W}_{cr}$ on strain rate. Zhao and co-workers \citep{fastdamagevolution1} synthesized a rate-dependent soft polymer with dynamic crosslinks [TFPM-PDMS-25000: 4-[tris(4-formylphenyl)methyl]benzaldehyde (TFPM) used as a tetratopic linker to crosslink PDMS]. The mechanical response of TFPM-PDMS-25000 under uniaxial tensile loading at different stretching speeds was also characterized in the paper. Results are shown in Figure \ref{Wcrexamples}A. We calculated $\mathit{W}_{cr}$ for each loading rate by evaluating the area under the corresponding engineering stress-strain curve. The variation of $\mathit{W}_{cr}$ with respect to strain rate is shown in Figure \ref{Wcrexamples}B. $\mathit{W}_{cr}$ is approximately constant over one decade of strain rate considered. Polyurethane's (2\% w/w of Polyethylene glycol with respect to Palm-based polyol content) rate-dependent mechanical response reported by Somarathna and co-workers \citep{Wcrexample2} is shown in Figure \ref{Wcrexamples}C with uniaxial tensile engineering stress-strain curves over a range of strain rates. We calculated $\mathit{W}_{cr}$ for each strain rate from the corresponding engineering stress-strain curve. We show that $\mathit{W}_{cr}$ is roughly constant over two decades of strain rate through Figure \ref{Wcrexamples}D. Over this range of strain rate, energy release rate, and crack tip field measures for rate-dependent soft polymers can vary by an order of magnitude or more \citep{Songfracture, Fracturechallenges1}. We also show a negligible variation in $\mathit{W}_{cr}$ value for different loading modes by evaluating $\mathit{W}_{cr}$ from the stress-strain data for Ecoflex elastomer (material has no loading mode-based response asymmetries) reported by Ahmad and co-workers \citep{AHMADdiffloadmodes}. Results of homogeneous uniaxial tension (UX), pure shear (PS), and equibiaxial tension (EB) tests performed by the authors at a constant engineering strain rate of 0.3 s$^{-1}$ are shown in Figure \ref{Wcrexamples}E. We evaluated $\mathit{W}_{cr}$ for each loading mode from the corresponding engineering stress-strain curves. $\mathit{W}_{cr}$ is approximately constant for the three loading modes considered: 2 MPa (UX), 1.6 MPa (PS), and 1.7 MPa (EB), as shown in Figure \ref{Wcrexamples}F. These examples further support our hypothesis that, within a reasonable approximation, $W_{cr}$ can be used as a scalar measure of fracture resistance of a certain class of soft polymers.

\section{Multi-mechanism fracture modeling framework}
\label{Multimechanism damage framework}

We follow the Kroner-Lee \citep{Kroner(1959), Lee(1969), Mjmetastablesteels, SijunplasticityPINN} multiplicative decomposition of the deformation gradient tensor ${F}_{{ij}}$ into elastic and plastic parts ${F}^e_{{ij}}$ and ${F}^p_{{ij}}$, respectively and the parallel multi-mechanism generalization of the multiplicative decomposition used in elastic-viscoplastic and viscoelastic constitutive theories for polymers \citep{BOYCE1988, BOYCE1989, Ekuhl1, Ekuhl2, BergstromBoyce(1998), AnandAmes(2009)I, Srivastavashapememory}. We consider an isothermal framework for materials that can be idealized as isotropic. For the general framework, $M$ parallel mechanisms of deformation are considered. Damage variables $d^{(q)}({X}_{i}, t) \in [0,1], \:\: q \in [1,M]$ corresponding to each micro-mechanism are introduced to model damage and subsequent fracture. We set the general criteria to be $d^{(q)}({X}_{i}, t)$ = 0 for undamaged and $d^{(q)}({X}_{i}, t)$ = 1 for completely damaged mechanism $q$ (when a mechanism $q$ reaches complete damage, it has no contribution to the total stress) at a material point ${X}_{i}$. $\mathit{W}_{cr}$ in a multi-mechanism sense can be interpreted as the value of $W$ when the damage fields for all mechanisms ($d^{(1)}, d^{(2)},...., d^{(M)}$) reach the value of 1 at a material point subjected to loading rates sufficiently fast to cause damage initiation in each mechanism. The method of virtual power is now used to derive the macroscopic and microscopic force balances.

\subsection{Method of virtual power for multi-mechanism deformation and damage}
\label{Method of virtual power}

The kinematic rate-like descriptors in the framework, while considering $M$ parallel mechanisms of deformation that capture the mechanical response of a material, are taken as 

\begin{equation}
    \dot{{\chi}}_{i}, \dot{{F}}_{ij}^{e(1)}, \dot{{F}}_{ij}^{e(2)}, ..., \dot{{F}}_{ij}^{e(M)}, \dot{\epsilon}^{p(1)}, \dot{\epsilon}^{p(2)}, ..., \dot{\epsilon}^{p(M)}, \dot{d}^{(1)}, \dot{d}^{(2)}, ..., \dot{d}^{(M)}, \dfrac{\partial \dot{d}^{(1)}}{\partial X_{i}}, \dfrac{\partial \dot{d}^{(2)}}{\partial X_{i}}, ..., \dfrac{\partial \dot{d}^{(M)}}{\partial X_{i}}.
\label{all rate like kinematic descriptors}
\end{equation}

\noindent ${{\chi}}_{i}$ is a smooth one-to-one mapping of a material point $X_{i}$ to its position in the deformed configuration $x_{i}$ as ${x}_{i} = {\chi}_{i}({X}_{1}, {X}_{2}, {X}_{3}, t)$. ${F}_{ij}=\dfrac{\partial \chi_{i}}{\partial X_{j}}$ is the deformation gradient tensor. Each mechanism at a material point is assumed to experience the same total deformation quantified by the deformation gradient tensor. ${F}_{ij}^{e(q)}$ and ${F}_{ij}^{p(q)}$ are the elastic and plastic parts of ${F}_{ij}$ in mechanism $q \in [1,M]$ following the multiplicative decomposition ${F}_{ij} = {F}_{ik}^{e(q)} {F}_{kj}^{p(q)}$.
 
 \begin{equation}
 \begin{split}
     &\dot{\epsilon}^{p(q)} \overset{\text{def}}{=} \sqrt{{D}_{ij}^{p(q)} {D}_{ij}^{p(q)}}, \: {D}_{ij}^{p(q)} = \dfrac{1}{2}\Big(\dot{{F}_{ik}}^{p(q)} {F}_{kj}^{p(q)^{-1}} + {F}_{ki}^{p(q)^{-1}} \dot{{F}_{jk}}^{p(q)}\Big), \: \epsilon^{p(q)} \overset{\text{def}}{=} \int_{0}^{t} \dot{\epsilon}^{p(q)} \: dt, \: \\& d^{(q)} \in [0,1] \:\: : q \in [1,M],
 \end{split}
\label{general measure of plastic strain rate}
 \end{equation}
 
\noindent are a measure of plastic strain rate, plastic stretching tensor corresponding to ${F}_{ij}^{p(q)}$, a measure of plastic strain, and damage variable, respectively, for mechanism $q$. The model is developed for isotropic materials, and consequently, the assumption of plastic irrotationality is valid. The definition of velocity gradient ${L}_{ij}$ (=$\dot{{F}}_{ik} {F}_{kj}^{-1}$) and the multiplicative decomposition of ${F}_{ij}$ result in the following constraints

\begin{equation}
    \dfrac{\partial \dot{{\chi}}_{i}}{\partial X_{k}} {F}_{kj}^{-1} = \dot{{F}}_{ik}^{e(q)} {F}_{kj}^{e(q)-1} + \dot{\epsilon}^{p(q)} {F}_{ik}^{e(q)} {N}_{kl}^{p(q)} {F}_{lj}^{e(q)-1}, \:\: {N}_{ij}^{p(q)} = \dfrac{{D}_{ij}^{p(q)}}{\Big(\sqrt{{D}_{kl}^{p(q)} {D}_{kl}^{p(q)}}\Big)} : \: \: \forall q \in [1,M],
\label{Kinematic constraint}    
\end{equation}

\noindent where ${N}_{ij}^{p(q)}$ is the plastic flow direction for mechanism $q$. Evolution of a reference body B and an arbitrary part of it denoted by P with outward unit normal ${n}^{\text{R}}_{i}$ on its boundary $\partial$P is associated with macroscopic and microscopic force systems. The macroscopic system is defined by (i) traction ${t}^{\text{R}}_{i}$ corresponding to ${n}^{\text{R}}_{i}$ and generalized body force per unit referential volume ${b}^{\text{R}}_{i}$ expending power over $\dot{{\chi}}_{i}$ on $\partial$P and in P, respectively. Specifically, ${b}^{\text{R}}_{i} = {b}^{0\text{R}}_{i} - \rho_{\text{R}} \Ddot{{\chi}}_{i}$,
with ${b}^{0\text{R}}_{i}$ being the conventional body force per unit referential volume, $- \rho_{\text{R}} \Ddot{{\chi}}_{i}$ representing the inertial body force per unit referential volume, $\rho_{\text{R}}$ denoting the referential mass density and (ii) elastic stress ${P}_{ij}^{e(q)}$ expending power over the elastic distortion rate $\dot{{F}}_{ij}^{e(q)}$ in P.

The microscopic system is defined with: (i) scalar microscopic stresses $\pi^{(q)}$ expending power over $\dot{\epsilon}^{p(q)}$ in P, (ii) scalar microscopic stresses $\overline{\omega}^{(q)}$ expending power over $\dot{d}^{(q)}$ in P, (iii) vector microscopic stresses $\xi_{i}^{(q)}$ expending power over $\dfrac{\partial \dot{d}^{(q)}}{\partial X_{i}}$ in P and (iv) scalar microscopic tractions $\Upsilon^{(q)} ({n}^{\text{R}}_{i})$ expending power over $\dot{d}^{(q)}$ on $\partial$P. For a given P, the power expended on P by the material external to it \Big($\mathcal{W}_{ext}$(P)\Big) and the power expenditure within P \Big($\mathcal{W}_{int}$(P)\Big) respectively, can be expressed as

\begin{equation}
\begin{split}
& \mathcal{W}_{ext}(\text{P}) = \int_{\partial \text{P}}{t}^{\text{R}}_{i} \dot{{\chi}}_{i} da_{\text{R}} + \int_{\text{P}}{b}^{\text{R}}_{i} \dot{{\chi}}_{i} dv_{\text{R}} + \int_{\partial \text{P}} \sum_{q=1}^{M} \Big({\Upsilon}^{(q)} \dot{d}^{(q)} \Big)\: da_{\text{R}}, \\&  \mathcal{W}_{int}(\text{P}) = \int_{\text{P}} \Big(\sum_{q=1}^{M} \Big[J^{p(q)} {P}_{ij}^{e(q)} \dot{{F}}_{ij}^{e(q)} + J^{p(q)} \pi^{(q)} \dot{\epsilon}^{p(q)} + \overline{\omega}^{(q)} \dot{d}^{(q)} + {\xi}_{i}^{(q)} \dfrac{\partial \dot{d}^{(q)}}{\partial X_{i}} \Big]\Big) dv_{\text{R}}.
\end{split}
\label{Wint and Wext definition}
\end{equation}

\noindent Note that the stress powers ${P}_{ij}^{e(q)} \dot{{F}}_{ij}^{e(q)}$, $\pi^{(q)} \dot{{\epsilon}}^{p(q)}$ in \eqref{Wint and Wext definition} are per unit volume ($J^{p(q)}$) of the intermediate space corresponding to ${F}_{ij}^{p(q)}$. Assuming that the fields ${\chi}_{i}, {F}_{ij}^{e(1)}, {F}_{ij}^{e(2)}, ..., {F}_{ij}^{e(M)}, \epsilon^{p(1)}, \epsilon^{p(2)}, \\..., \epsilon^{p(M)}, d^{(1)}, d^{(2)}, ..., d^{(M)}$ are known at an arbitrarily chosen time instant, the fields $\dot{{\chi}_{i}}, \dot{{F}}_{ij}^{e(1)}, \dot{{F}}_{ij}^{e(2)},\\ ..., \dot{{F}}_{ij}^{e(M)},  \dot{\epsilon}^{p(1)}, \dot{\epsilon}^{p(2)}, ..., \dot{\epsilon}^{p(M)}, \dot{d}^{(1)}, \dot{d}^{(2)}, ..., \dot{d}^{(M)}$ need to be specified independently while satisfying the constraints in \eqref{Kinematic constraint}. The generalized virtual velocity can be written as 

\begin{equation}
    {\mathcal{V}} = ({\tilde{{\chi}}}_{i}, \tilde{{F}}_{ij}^{e(1)}, \tilde{{F}}_{ij}^{e(2)}, ..., \tilde{{F}}_{ij}^{e(M)}, \tilde{{\epsilon}}^{p(1)}, \tilde{{\epsilon}}^{p(2)}, ..., \tilde{{\epsilon}}^{p(M)}, \tilde{d}^{(1)}, \tilde{d}^{(2)}, ..., \tilde{d}^{(M)}).
\label{virtual velocity defintion}
\end{equation} 

\noindent The virtual field ${\mathcal{V}}$ is rigid if it satisfies 

\begin{equation}
    (\nabla \tilde{{\chi}}_{i}) = \tilde{{F}}_{ij} = {\Omega}_{ik} {F}_{kj}, \:\: \tilde{{F}}_{ij}^{e(q)} = {\Omega}_{ik} {F}_{kj}^{e(q)}, \: \tilde{{\epsilon}}^{p(q)} = 0, \: \tilde{d}^{(q)} = 0 : \forall q \in [1,M],
\label{rigid virtual velocity expression 1}
\end{equation}

\noindent where ${\Omega}_{ij}$ is a spatially constant skew tensor. Here, $\tilde{(.)}$ denotes virtual field quantities. The principle of virtual power requires that 

\begin{equation}
    \mathcal{W}_{int}(\text{P}, {\mathcal{V}}) = \mathcal{W}_{ext}(\text{P}, {\mathcal{V}}) \: : \forall {\mathcal{V}} \: \text{and} \:\: \mathcal{W}_{int}(\text{P}, {\mathcal{V}}) = 0 \: : \forall \: \text{rigid} \: {\mathcal{V}}.
\label{principle of virtual power 1}
\end{equation}

The total free energy density per unit referential volume $\psi_{\text{R}}$ is assumed to be additively decomposable into contributions from each mechanism of deformation and energetic parts accounting for the effects of gradients in the damage fields $\psi^{d(q)}$: $q \in [1, M]$ as 

\begin{equation}
\psi_{\text{R}} = \sum_{q=1}^{M} \Big(J^{p(q)} \psi^{(q)} + \psi^{d(q)}\Big), \: \: {T}_{ij} = \sum_{q=1}^{M} {T}_{ij}^{(q)}, 
\label{general additive decomposition of psiR and Cauchy stress}
\end{equation}

\noindent where $\psi^{(q)}$ is the contribution from mechanism $q$ per unit volume of the corresponding intermediate space ($J^{p(q)}$). ${T}_{ij}, {T}_{ij}^{(q)}$ are the total Cauchy stress and total Cauchy stress contribution from mechanism $q$, respectively. $\psi^{d(q)}$ can be interpreted as defect energy due to local elastic incompatibilities caused by gradients in the damage field variable $d^{(q)}$. This is similar to the interpretation of the free energy contributions due to plastic strain gradients in standard strain-gradient plasticity theories \citep{gradientphysintpr1, gradientphysintpr2, GAOstraingradientplasticity}. The total first Piola-Kirchoff stress ${P}_{ij}$, the total second Piola-Kirchoff stress ${S}_{ij}$ and ${T}_{ij}$ are related through the following relations 

\begin{equation}
\begin{split}
    &{P}_{ij} = J {T}_{ik} {F}_{jk}^{-1} = J \Big(\sum_{q=1}^{M} {T}_{ik}^{(q)}\Big) {F}_{jk}^{-1} = \sum_{q=1}^{M} J {T}_{ik}^{(q)} {F}_{jk}^{-1} = \sum_{q=1}^{M} {P}_{ij}^{(q)}, \: {P}_{ij} = {F}_{ik} {S}_{kj}, \\& {S}_{ij} = J {F}_{ik}^{-1} {T}_{kl} {F}_{jl}^{-1},
\end{split}    
\label{First PK stress relation with Cauchy and Second PK stress}
\end{equation}

\noindent where ${P}_{ij}^{(q)}$ is the total first Piola-Kirchoff stress contribution from mechanism $q$.

\subsection{Consequences of the method of virtual power: Macroscopic and microscopic stresses for multi-mechanism deformation and damage}
\label{PVP consequences}

The principle of virtual power necessitates that \eqref{principle of virtual power 1} is satisfied. To deduce the consequences, any generalized virtual velocity ${\mathcal{V}}$ consistent with \eqref{Kinematic constraint} can be chosen. Standard procedure and variational arguments to obtain macroscopic and microscopic force balances \citep{PVP1, PVP2, PVP3, Gurtinfriedanand, Anandquasibrittle} are followed, and the results are summarized.

\subsubsection{Macroscopic stresses} 

A ${\mathcal{V}}$ with $\tilde{\epsilon}^{p(q)} \equiv 0, \: \tilde{d}^{(q)} \equiv 0 \: : \forall q \in [1,M]$ is considered, so that $ \dfrac{\partial \tilde{\chi_{i}}}{\partial X_{j}} = \tilde{F}_{ik}^{e(q)} {F}_{kj}^{p(q)}$. Using \eqref{First PK stress relation with Cauchy and Second PK stress} and standard arguments, \eqref{principle of virtual power 1} results in

\begin{equation}
    {t}^{\text{R}}_{i} = {P}_{ij} {n}^{\text{R}}_{j}, \:\:  \dfrac{\partial {P}_{ij}}{\partial X_{j}} + {b}^{0\text{R}}_{i} = \rho_{\text{R}} \ddot{{\chi}}_{i} : \:\: {P}_{ij}^{(q)} \overset{\text{def}}{=} J^{p(q)} {P}_{ik}^{e(q)} {{F}}_{jk}^{p(q)-1} \: \forall q \in [1,M].
\label{conventional traction condition and force balance}
\end{equation}

\noindent \eqref{conventional traction condition and force balance} represents the standard referential forms of macroscopic traction condition and force balance, respectively. The second requirement of the principle of virtual power, as given in \eqref{principle of virtual power 1}, leads to 

\begin{equation}
    {P}_{ik} {F}_{jk} = {F}_{ik} {P}_{jk},
\end{equation}

\noindent which is the standard result obtained from macroscopic moment balance.

\subsubsection{Microscopic stresses}

(i) Scalar microstresses $\pi^{(q)}$

\noindent Next we consider a ${\mathcal{V}}$ with $\tilde{{\chi}}_{i} \equiv 0_{i}$, $\tilde{d}^{(q)} \equiv 0 \: : \forall q \in [1,M]$ and arbitrarily chosen $\tilde{\epsilon}^{p(q)} \: : q \in [1,M]$. Two additional stress measures are defined, (i) Symmetric elastic second Piola-Kirchoff stress and (ii) Mandel stress, respectively, as

\begin{equation}
    {S}_{ij}^{e(q)} \overset{\text{def}}{=} J^{e(q)} {F}_{ik}^{e(q)-1} {T}_{kl}^{(q)} {F}_{jl}^{e(q)-1}, \:\: {M}_{ij}^{e(q)} \overset{\text{def}}{=}  {C}_{ik}^{e(q)} {S}_{kj}^{e(q)} = J^{e(q)} {F}_{ki}^{e(q)} {T}_{kl}^{(q)} {F}_{jl}^{e(q)-1} \: : \forall q \in [1,M].
\label{elastic second PK stress definition}
\end{equation}

\noindent $C_{ij}^{e(q)} = F_{ki}^{e(q)} F_{kj}^{e(q)}$ is the elastic right Cauchy-Green deformation tensor for mechanism $q$. The first requirement of the principle of virtual power given in \eqref{principle of virtual power 1} results in the microscopic force balance

\begin{equation}
    \pi^{(q)} = \dfrac{1}{2}\Big({M}_{ij}^{e(q)} + {M}_{ji}^{e(q)}\Big) {N}_{ij}^{p(q)} \:\:\: \forall q \in [1,M].
\label{final result 1st requirement PVP pi}
\end{equation}

(ii) Scalar microstresses $\overline{\omega}^{(q)}$ and vector microstresses ${\xi}_{i}^{(q)}$

\noindent To obtain $\overline{\omega}^{(q)}$ and ${\xi}_{i}^{(q)}$, a ${\mathcal{V}}$ with $\tilde{\epsilon}^{p(q)} \equiv 0 \: : \forall q \in [1,M]$ and $\tilde{{\chi}}_{i} \equiv 0_{i}$ is considered. Choosing $\tilde{d}^{(q)} \: : q \in [1,M] $ arbitrarily, \eqref{principle of virtual power 1} yields the microscopic traction conditions and force balances respectively as

\begin{equation}
    \Upsilon^{(q)} = {\xi}_{i}^{(q)} {n}^{\text{R}}_{i}, \:\: \dfrac{\partial {\xi}_{i}^{(q)}}{\partial X_{i}} - \overline{\omega}^{(q)} = 0 \: : \forall q \in [1,M].
\label{microscopic traction condition final xi}
\end{equation}

\noindent A set of thermodynamically consistent constitutive equations, along with the above macroscopic and microscopic force balances, give the complete set of governing mechanical equations for the model. The internal power expenditure in \eqref{Wint and Wext definition} can now be rewritten using \eqref{elastic second PK stress definition} as 

\begin{equation}
   \mathcal{W}_{int}(\text{P}) = \int_{\text{P}} \Big (\sum_{q=1}^{M} \Big[\frac{1}{2} J^{p(q)} {S}_{ij}^{e(q)} \dot{{C}}_{ij}^{e(q)} + J^{p(q)} \pi^{(q)} {\dot{\epsilon}}^{p(q)} + \overline{\omega}^{(q)} \dot{d}^{(q)} + {\xi}_{i}^{(q)} \dfrac{\partial \dot{d}^{(q)}}{\partial X_{i}}\Big]\Big) dv_{\text{R}}. 
\label{final Wint expression}
\end{equation}

\subsection{Free energy imbalance}
\label{Free energy imbalance}

Under isothermal conditions, the local form of free energy imbalance can be finally expressed as

\begin{equation}
    \dot{\psi}_{\text{R}} - \sum_{q=1}^{M} \Big(\frac{1}{2} J^{p(q)} {S}_{ij}^{e(q)} \dot{{C}}_{ij}^{e(q)} + J^{p(q)} \pi^{(q)} {\dot{\epsilon}}^{p(q)} + \overline{\omega}^{(q)} \dot{d}^{(q)} + {\xi}_{i}^{(q)} \dfrac{\partial \dot{d}^{(q)}}{\partial X_{i}}\Big) \leq 0.
\label{local form free energy imbalance}    
\end{equation}

\noindent We note here that all the terms involved in \eqref{local form free energy imbalance} are frame invariant.

\subsection{Constitutive equations for deformation and damage}
\label{Constitutive equations}

Using \eqref{general additive decomposition of psiR and Cauchy stress}, the free energy imbalance in \eqref{local form free energy imbalance} can be written as

\begin{equation}
\begin{split}
    &\sum_{q=1}^{M} \Bigg(J^{p(q)} \Big(\dfrac{\partial \hat{\psi}^{(q)}({\Lambda}^{(q)})}{\partial {C}_{ij}^{e(q)}} - \frac{1}{2}  {{S}}_{ij}^{e(q)}\Big) \dot{{C}}_{ij}^{e(q)}  - J^{p(q)} \Big(\pi^{(q)} - \dfrac{\partial \hat{\psi}^{(q)}({\Lambda}^{(q)})}{\partial \epsilon^{p(q)}}\Big) \dot{\epsilon}^{p(q)} - \Big(\overline{\omega}^{(q)}-\dfrac{\partial \tilde{\psi}_{\text{R}}({\Lambda})}{\partial d^{(q)}}\Big) \dot{d}^{(q)} \\& - \Big({\xi}_{i}^{(q)} - \dfrac{\partial \tilde{\psi}_{\text{R}}({\Lambda})}{\partial (\dfrac{\partial {d}^{(q)}}{\partial X_{i}})}\Big) \dfrac{\partial \dot{d}^{(q)}}{\partial X_{i}} \Bigg) \leq 0.
\end{split}    
\label{plugging constitutive relations in free energy imbalance}
\end{equation}

\noindent where

\begin{equation}
    {\Lambda}^{(q)} = \Big({C}_{ij}^{e(q)}, \epsilon^{p(q)}, d^{(q)}, \dfrac{\partial {d}^{(q)}}{\partial X_{i}}\Big) \: \: q \in [1,M], \: {\Lambda} = ({\Lambda}^{(1)}, {\Lambda}^{(2)}, ..., {\Lambda}^{(M)}),
\label{psii general dependence}
\end{equation}

\noindent are generalized lists of variables. $\psi^{d(q)}$ is taken to be dependent on $d^{(q)}$, $\dfrac{\partial {d}^{(q)}}{\partial X_{i}}$. The following state relations are assumed

\begin{equation}
    {S}_{ij}^{e(q)} = 2 \dfrac{\partial \hat{\psi}^{(q)}({\Lambda}^{(q)})}{\partial {C}_{ij}^{e(q)}}, \:\: {\xi}_{i}^{(q)} = \dfrac{\partial \tilde{\psi}_{\text{R}}({\Lambda})}{\partial (\dfrac{\partial {d}^{(q)}}{\partial X_{i}})}, \:\: \pi^{(q)}_{en} = \dfrac{\partial \hat{\psi}^{(q)}({\Lambda}^{(q)})}{\partial \epsilon^{p(q)}}, \:\: \overline{\omega}^{(q)}_{en} = \dfrac{\partial \tilde{\psi}_{\text{R}}({\Lambda})}{\partial d^{(q)}} \: : \forall q \in [1,M],
\label{state relation 2nd PK stress and vector microstress}
\end{equation}

\noindent with $\pi^{(q)} = \pi^{(q)}_{en} + \pi^{(q)}_{dis}$, $\overline{\omega}^{(q)} = \overline{\omega}^{(q)}_{en} + \overline{\omega}^{(q)}_{dis}$. The subscripts $en$ and $dis$ denote energetic and dissipative parts, respectively. To satisfy the dissipation inequality in \eqref{plugging constitutive relations in free energy imbalance}, it is assumed that 

\begin{equation}
     \pi^{(q)}_{dis} \dot{\epsilon}^{p(q)} \geq 0, \:\: \overline{\omega}^{(q)}_{dis} \dot{d}^{(q)} \geq 0 \:\: \forall q \in [1,M].
\label{individual term dissipation inequality}
\end{equation}

The constitutive dependence of ${\psi}_{\text{R}}, {\psi}^{(q)}$, and ${\psi}^{d(q)}$ on the generalized lists of variables in \eqref{psii general dependence} can be simplified without significant loss of generality. Specifically,

\begin{equation}
    \psi_{\text{R}} = \sum_{q=1}^{M}\Big[g^{(q)}(d^{(q)})\psi^{(q)+}_{\text{o}} + \psi^{(q)-}_{\text{o}} + \underbrace{\psi^{(q)}_{*} \: [l^{(q)}]^{2} \Big(\dfrac{\partial d^{(q)}}{\partial X_{i}} \dfrac{\partial d^{(q)}}{\partial X_{i}}\Big)}_{\psi^{d^{(q)}}}\Big], \:\: \psi_{\text{o}} = \sum_{q=1}^{M} \psi^{(q)}_{\text{o}}({C}_{ij}, {C}_{ij}^{e(q)}), 
\label{psiR function minimized}
\end{equation}

\noindent where $ 0 \le g^{(q)}(d^{(q)}) \le$ 1 are monotonically decreasing functions which degrade the energy storage with damage growth at a material point. $\psi^{(q)}_{*}>0$ and $l^{(q)} >$ 0 are coefficients with units of energy per unit volume and length scale parameters that control the spread of the diffuse damage zone corresponding to $d^{(q)}$, respectively. $\psi_{\text{o}}$ is the total undamaged elastic free energy, which is the additive decomposition of contributions ($\psi^{(q)}_\text{o}$) from the $M$ mechanisms of deformation. $\psi^{(q)-}_{\text{o}}$ is the part of $\psi^{(q)}_{\text{o}}$ which is dilatational and compressive. $\psi^{(q)+}_{\text{o}} = \psi^{(q)}_{\text{o}} - \psi^{(q)-}_{\text{o}}$ is the damage driving part of $\psi_{\text{o}}^{(q)}$. This split accounts for damage growth being driven only by distortional and tensile dilatational deformation \citep{HDsplit1, HDsplit2}.\\

Further, using \eqref{psiR function minimized} and \eqref{state relation 2nd PK stress and vector microstress}, we get

\begin{equation}
    \pi^{(q)}_{en} = 0, \:\: \overline{\omega}^{(q)}_{en} = \dfrac{\partial g^{(q)}(d^{(q)})}{\partial d^{(q)}} \psi^{(q)}_{\text{o}}, \:\: {\xi}_{i}^{(q)} = 2\psi^{(q)}_{*}[l^{(q)}]^{2} \dfrac{\partial d^{(q)}}{\partial X_{i}}.
\label{energetic stresses minimized}
\end{equation}

\noindent Also, $\pi^{(q)}_{dis}$, $\overline{\omega}^{(q)}_{dis}$ can be taken in their simplified general forms as

\begin{equation}
    \pi^{(q)}_{dis} = g^{(q)}(d^{(q)}) \mathcal{S}^{(q)}({C}_{ij}^{e(q)}, \dot{\epsilon}^{p(q)}) \:\: \text{when} \:\: \dot{\epsilon}^{p(q)} > 0, \:\: \overline{\omega}^{(q)}_{dis} = \hat{\overline{\omega}}^{(q)}_{dis}(d^{(q)}, \dot{d}^{(q)}), \:\: \text{when} \:\: \dot{d}^{(q)} > 0
\label{dissipative stresses minimized}    
\end{equation} 

\noindent with $\mathcal{S}^{(q)}({C}_{ij}^{e(q)}, \dot{\epsilon}^{p(q)})$, $\hat{\overline{\omega}}^{(q)}_{dis}(d^{(q)}, \dot{d}^{(q)})$ being non-negative so that the reduced dissipation inequalities in \eqref{individual term dissipation inequality} are satisfied.

\section{Framework specialization to model soft polymers}

The functional forms for the deformation-related quantities during specialization, specifically $\psi_{\text{o}}$ and $\textbf{T}_{\text{o}}$, are taken from our previous work \citep{KonalePBS}. In the following sub-sections, the multi-mechanism fracture modeling framework developed in Section \ref{Multimechanism damage framework} is specialized along with the $\psi_{\text{o}}$, $\textbf{T}_{\text{o}}$ expressions to model damage initiation and evolution. $\psi_{\text{o}}^{+}$ is introduced and discussed as a damage driving force with $\psi_{cr}^{+}$ (the energetic part of $W_{cr}$) as a damage initiation criterion.

\subsection{Free energy and Cauchy stress}
\label{Free energy}

To specialize the fracture modeling framework for soft polymers, we restrict the number of micro-mechanisms $M$ to 2 in \eqref{psiR function minimized}. As an approximation, only one damage variable is used. This reflects the commonly observed single load drop corresponding to fracture in experiments for a broad class of soft polymers in the literature. The single load drop is also shown in our experiments on notched PBS specimens (Figure \ref{Calibration}B). The resistance to deformation results from the underlying bonds and their networks. For a material point to be completely damaged, both mechanisms $A$ and $B$ have to be completely damaged, i.e., all bonds resisting deformation must be broken. In the case of multiple load drop observation \citep{Multiloaddrop1, Multiloaddrop2}, different damage variables for each mechanism need to be used. Some works on fracture modeling of composites have incorporated multiple damage variables to account for matrix and filler damage separately \citep{Doublepf1, Doublepf2, Weican2024}. 

\subsubsection{Constitutive model for viscous soft polymer PBS}

The two mechanisms of deformation considered are: (a) \textbf{Intermolecular Resistance} ($A$): Elastic and dissipative resistance due to intermolecular interactions between neighboring molecules \citep{AMES2009, Srivastava2011Thermo, Lamont2025}. (b) \textbf{Network Resistance} ($B$): stretching of constituent subchains within Boron(B):Oxygen(O) coordinate-bond dynamic crosslinks \citep{KonalePBS}. 

$\psi_{\text{o}}$ is given as 

\begin{equation}
\begin{split}
\psi_{\text{o}} &= \underbrace{{\psi}_{\text{o}}^{(A)}}_\text{Intermolecular resistance} + \underbrace{{\psi}_{\text{o}}^{(B)}}_\text{Network resistance}, \\& = \underbrace{G \: \Big({E}_{ij}^{e(A)`} {E}_{ij}^{e(A)`}\Big) +  \frac{1}{2} K \left({E}_{kk}^{e(A)}\right)^2}_{\psi^{(A)}_{\text{o}}} \\& +\underbrace{\mu \: \overline{\lambda}^{2}_L \Big(e^{-k_{ns} t} \: \zeta(t/0) + \int_{0}^{t}  \dfrac{n_{ns}}{n_s}  k_{s} e^{-k_{ns}(t-t^{'})} \: \zeta(t/t^{'}) \: dt^{'}\Big)}_{\psi^{(B)}_{\text{o}}}, \\& \text{with} \:\:\: \zeta(t/t^{'}) = \left[ \left(\dfrac{\overline{\lambda}(t/t^{'})}{\overline{\lambda}_L}\right)\beta(t/t^{'}) + ln\left(\dfrac{\beta(t/t^{'})}{\text{sinh}\beta(t/t^{'})}\right)\right].
\\& \text{if} \: {E}_{kk}^{e(A)} < 0: {\psi}^{(A)+}_{\text{o}} = G \: \Big({E}_{ij}^{e(A)`} {E}_{ij}^{e(A)`}\Big),\:\:{\psi}^{(A)-}_{\text{o}} = \dfrac{1}{2} K \left({E}_{kk}^{e(A)}\right)^2, \\& \text{else}: \: {\psi}^{(A)+}_{\text{o}} = {\psi}^{(A)}_{\text{o}},  \:\: {\psi}^{(A)-}_{\text{o}} = 0.
\label{undamaged free energy expression}
\end{split}
\end{equation}

\noindent We note that $\psi^{(B)}_{\text{o}}$ is purely deviatoric. Any suitable free energy density base function can be used in \eqref{undamaged free energy expression} for a specific soft polymer under consideration. $G>0$, $K>0$ are the shear and bulk moduli respectively. ${E}^{e(A)}_{ij}=\mathrm{ln} {U}_{ij}^{e(A)}$ is an elastic logarithmic strain measure, where ${U}_{ij}^{e(A)}$ is the right elastic stretch tensor following the polar decomposition of ${F}_{ij}^{e(A)}$. ${A}^{`}_{ij}$ denotes the deviatoric part of a tensor ${A}_{ij}$. Plastic incompressibility is assumed for mechanism $A$. $\overline{\lambda}_L$ is the network locking stretch parameter accounting for finite chain extensibility. $\overline{\lambda} \overset{\text{def}}{=} \sqrt{\dfrac{{C}^{*}_{kk}}{3}}$ is the effective total distortional stretch with ${C}^{*}_{ij}$ being the distortional part of ${C}_{ij}$. $\mathcal{L}(.) = \text{coth}(.) - \dfrac{1}{(.)}$ denotes the Langevin function and $\beta = \mathcal{L}^{-1}\left(\dfrac{\overline{\lambda}}{\overline{\lambda}_L}\right)$. The referential number density of chain segments currently in their subchain state (S) and that of chain segments that can be in the subchain state but are not currently (NS) are denoted by $n_s$ and $n_{ns}$, respectively. The underlying dynamic, reversible reactions can be represented as

\begin{equation}
   n_{ns} \:\:\:\: (\mathrm{NS}) \underset{k_{ns}}{\stackrel{k_{s}}{\rightleftharpoons}} (\mathrm{S}) \:\:\: n_{s}, 
   \label{generalized reversible crosslink reaction}
\end{equation}

\noindent where $k_{s}$ and $k_{ns}$ are the forward and backward reaction rate parameters, respectively. $\mu = n_{s} k_{\text{B}} \theta$ denotes the ground state shear modulus with $k_{\text{B}}$, $\theta$ denoting the Boltzmann constant and temperature respectively. Kinematic quantities with notation ($t/t^{'}$) use the relative deformation gradient: ${F}_{ij}(t/t^{'}) = {F}_{ik}(t) {F}_{kj}^{-1}(t^{'})$, a temporal decomposition of ${F}_{ij}$, which maps the deformed configuration at $t^{'}$ to the deformed configuration at $t$ as the input. This accounts for subchains formed at $t^{'}$ ($0 < t^{'} < t$) and surviving till the current time $t$ being stretched with respect to their stress-free, relaxed configurations corresponding to $t^{'}$. ${\psi}_{\text{o}}^{(B)}$ in \eqref{undamaged free energy expression} is the summation of the stretching energies of these subchains with different stretch levels corresponding to their formation time instants. This proposed free energy formulation avoids having a thermodynamically inconsistent, explicitly rate-dependent free energy density function that is sometimes proposed in the literature. The degradation function $g(d)$ is taken as \citep{MIEHEgd, Maochainscission, ANANDgdHembtrl, Shabnamphasefield, Anand(2021)flowrule}

\begin{equation}
    g(d) = (1-d)^{2}.
\label{g(d) definition}
\end{equation}

\noindent The total Cauchy stress is given as

\begin{equation}
\begin{split}
&{T}_{ij} = g(d)\Big[\underbrace{{T}^{(A)\text{o}+}_{ij} + {T}^{(B)\text{o}}_{ij}}_{{T}^{\text{o}+}_{ij}}\Big] + \underbrace{{T}^{(A)\text{o}-}_{ij}}_{{T}^{\text{o}-}_{ij}}, \\& \text{if} \: {E}_{mm}^{e(A)} < 0: {T}^{(A)\text{o}+}_{ij} = J^{-1} {R}_{ik}^{e(A)} \Big(2 G {E}_{kl}^{e(A)`} \Big) {R}_{jl}^{e(A)}, \\& \:\:\:\:\:\:\:\:\:\:\:\:\:\:\:\:\:\:\:\:\:\:\:\:\:\:\: {T}^{(A)\text{o}-}_{ij} = J^{-1} {R}_{ik}^{e(A)} \Big(K  {E}_{mm}^{e(A)}  \delta_{kl} \Big) {R}_{jl}^{e(A)}, \\& \text{else}: \: {T}^{(A)\text{o}+}_{ij} = J^{-1} {R}_{ik}^{e(A)} \Big(2 G {E}_{kl}^{e(A)`} + K  {E}_{mm}^{e(A)}  \delta_{kl} \Big) {R}_{jl}^{e(A)}, \\& \:\:\:\:\:\:\:\:\:\:\:\: {T}^{(A)\text{o}-}_{ij} = 0, \\& {T}^{(B)\text{o}}_{ij} = e^{-k_{ns} t} \mu {\Gamma}_{ij}(t/0) +  \int_{0}^{t} \dfrac{n_{ns}}{n_s} k_s e^{-k_{ns} (t-t^{'})} \mu \: {\Gamma}_{ij}(t/t^{'}) \: dt^{'}, \\&{\Gamma}_{ij}(t/t^{'}) = J^{-1}(t/t^{'}) \dfrac{\overline{\lambda}_L}{\overline{\lambda}(t/t^{'})} \mathcal{L}^{-1}\Bigl[\dfrac{\overline{\lambda}(t/t^{'})}{\overline{\lambda}_L}\Bigr] {B}^{*`}_{ij}(t/t^{'}). 
\end{split}
\label{Total Cauchy stress final expression}
\end{equation}

\noindent ${T}^{\text{o}}_{ij}$ = ${T}^{\text{o}+}_{ij}$ + ${T}^{\text{o}-}_{ij}$ is the undamaged total Cauchy stress. ${T}^{\text{o}+}_{ij}$, ${T}^{\text{o}-}_{ij}$ are the undamaged total Cauchy stress contributions corresponding to ${\psi}_{\text{o}}^{+}$, ${\psi}_{\text{o}}^{-}$ respectively. ${T}^{(A)\text{o}}_{ij}$ = ${T}^{(A)\text{o}+}_{ij}$ + ${T}^{(A)\text{o}-}_{ij}$ and ${T}^{(B)\text{o}}_{ij}$ are the undamaged total Cauchy stress contributions from mechanisms $A$ \Big(${\psi}^{(A)}_{\text{o}} = {\psi}^{(A)+}_{\text{o}} + {\psi}^{(A)-}_{\text{o}}$\Big) and $B$ (${\psi}^{(B)}_{\text{o}}$) respectively \citep{KonalePBS}. ${B}_{ij} = {F}_{ik} {F}_{jk}$ is the total left Cauchy-Green deformation tensor.

\subsubsection{Broader applicability of the general constitutive model for PBS}

Setting $G=k_{ns}=k_{s} = 0$ in \eqref{undamaged free energy expression} results in $\psi_{\text{o}} = \mu \: \overline{\lambda}^{2}_L \: \zeta(t/0)$, which is the eight-chain free energy density function \citep{ArrudaBoyce(1993)} used for elastomers. The effect of $k_{ns}$'s value on the mechanical response can be understood easily by setting $G= 0$ in \eqref{undamaged free energy expression} to get $\psi_{\text{o}} = \psi^{(B)}_{\text{o}}$. At any given time instant, smaller values of $k_{ns}$ will result in $\psi_{\text{o}}$ tending to the eight-chain form $\mu \: \overline{\lambda}^{2}_L \: \zeta(t/0)$, i.e., elastomer-like behavior with insignificant network reconfiguration. On the other hand, higher values of $k_{ns}$ will lead to the departure of $\psi_{\text{o}}$ from the elastic behavior to significantly viscous behavior with appreciable network reconfiguration. Hence, \emph{the proposed constitutive model can be used as a general model for a broad class of soft polymers with response varying from elastic to significantly viscous.}

\subsection{Viscous flow rule for Mechanism $A$}
\label{Flow rule mechanism A m}
\vspace{0.05in}

Assuming plastic irrotationality, plastic incompressibility, and co-directionality of plastic flow, the plastic stretching tensor ${D}_{ij}^{p(A)}$ can be expressed as

\begin{equation}
	{D}_{ij}^{p(A)}= \nu^{p(A)} \left(\frac{{M}_{ij}^{e(A)`}}{2 \overline{\tau}^{(A)}} \right), \hspace{0.1in} \nu^{p(A)}\overset{\text{def}}{=}\sqrt{2} \sqrt{{D}_{ij}^{p(A)} {D}_{ij}^{p(A)}}, \hspace{0.1in} \overline{\tau}^{(A)}\overset{\text{def}}{=}\dfrac{1}{\sqrt{2}} \sqrt{{M}_{ij}^{e(A)`} {M}_{ij}^{e(A)`}},
    \label{Plastic stretching tensor, equivalent plastic shear strain rate (Mechanism A)}
\end{equation}

\noindent where $\nu^{p(A)}$, $\overline{\tau}^{(A)}$ are the equivalent plastic shear strain rate and the equivalent shear stress, respectively. With $\gamma^{p(A)} \overset{\text{def}}{=} \int_{0}^{t} \nu^{p(A)} \: dt$ being an equivalent plastic shear strain rate, we have $\nu^{p(A)} = \sqrt{2} \dot{\epsilon}^{p(A)}, \:\: \gamma^{p(A)} = \sqrt{2} \epsilon^{p(A)}$. Instead of $\dot{\epsilon}^{p(A)}$, we use $\nu^{p(A)}$ as the rate-like kinematic descriptor. The expression for $\pi^{(A)}$ in \eqref{final result 1st requirement PVP pi} will have to be divided by $\sqrt{2}$ to account for the change of $\dot{\epsilon}^{p(A)}$ to $\nu^{p(A)}$ as the kinematic descriptor. Using \eqref{final result 1st requirement PVP pi}, \eqref{energetic stresses minimized} and \eqref{Plastic stretching tensor, equivalent plastic shear strain rate (Mechanism A)}, we have $\overline{\tau}^{(A)} = \pi^{(A)} = \pi_{dis}^{(A)}$. 

Further, it is assumed that $\mathcal{S}^{(A)}$ in \eqref{dissipative stresses minimized} depends on ${C}_{ij}^{e(A)}$ only through $\hat{p}^{(A)} {=} - K {E}_{kk}^{e(A)} \\= - K \Big(\dfrac{1}{2}\mathrm{trace}(\text{ln} {C}_{ij}^{e(A)})\Big)$. Also, $\hat{p}^{(A)} = \dfrac{\overline{p}^{(A)}}{g(d)}$ with $\overline{p}^{(A)} \overset{\text{def}}{=} -\dfrac{1}{3} {M}_{kk}^{e(A)}$. $\mathcal{S}^{(A)}({C}_{ij}^{e(A)}, \nu^{p(A)})$ is chosen to follow a specific form as

\begin{equation}
    \mathcal{S}^{(A)}({C}_{ij}^{e(A)}, \nu^{p(A)}) = \alpha_{p} \hat{p}^{(A)} + \mathsf{S} \: \text{sinh}^{-1} \Bigg[\Bigg(\dfrac{\nu^{p(A)}}{\nu_{0}}\Bigg)^{m}\Bigg],
\label{S final expression}
\end{equation}

\noindent where $\alpha_{p} \geq 0$ is a pressure sensitivity parameter. The last term in \eqref{S final expression} represents a rate-dependent resistance to plastic flow. $\mathsf{S}$ is an internal variable that models the dissipative friction-like resistance due to the sliding chains. $\nu_{0} > 0$ is a reference strain rate and $m \in (0,1$] is a strain rate sensitivity parameter. The evolution equation for $\mathsf{S}$ is taken as $\dot{\mathsf{S}}=h \left(\overline{\lambda} -1 \right) \nu$  with initial value $\mathsf{S}({X}_{i},0)=\mathsf{S}_0 > 0$. $h$ is a stress-dimensioned material parameter and $\nu \overset{\text{def}}{=} \sqrt{2} \sqrt{{D}^{`}_{ij} {D}^{`}_{ij}}$ is the total equivalent shear strain rate. The strength relation in \eqref{dissipative stresses minimized} for mechanism $A$ can now be concisely expressed as 

\begin{equation}
	 \nu^{p(A)}=\begin{cases}
\begin{aligned}
			&0 \hspace{0.2in} \hspace{0.1in} &&\text{if}  \hspace{0.1in}\overline{\tau}^{(A)}_e \leq 0,\\
			&\nu_0 \left[\mathrm{sinh} \left({\frac{\overline{\tau}^{(A)}_e}{g(d)\mathsf{S}}} \right) \right] ^{1/{m}}  && \text{if} \hspace{0.1in} \overline{\tau}^{(A)}_e>0,
\end{aligned}
	\end{cases}
\label{Final overstress equivalent plastic shear strain rate relation}
\end{equation}

\noindent where $\overline{\tau}^{(A)}_e \overset{\text{def}}{=} \overline{\tau}^{(A)} - \alpha_{p} \overline{p}^{(A)}$ denotes a net equivalent shear stress for plastic flow.

\subsection{Evolution equation for damage variable $d$}
\label{Evolution equation for d m}

The microstructural changes leading to damage in soft polymers like PBS that exhibit dynamic crosslinking are thought to be reversible, i.e., failure of weak dynamic, reversible crosslinks \citep{Songfracture, Shenmodeling, Shenmodeling2}. However, \emph{short timescales for complete damage growth} with respect to the dynamic crosslink macroscopic formation timescale ($\tau_f$) would suggest that damage, if initiated, will grow monotonically. Specimen geometry and experiment schematic for a notched tension test performed at 60 mm/s loading speed are shown in Figure \ref{Calibration}(b). The sharp load drop corresponding to fracture in Figure \ref{Calibration}(b) for the notched PBS specimen \Big($\tau_f$ is in the order of seconds for PBS \citep{KonalePBS}\Big) highlights the rapid damage growth in PBS. Therefore, we adhere to the modeling approach for irreversible damage growth \citep{Anandquasibrittle, Shabnamphasefield}. 

Following Narayan and Anand \citep{Anandquasibrittle}, the scalar microstress $\overline{\omega}_{dis}$ in \eqref{dissipative stresses minimized} is taken in the following specific form 

\begin{equation}
    \overline{\omega}_{dis} = \alpha + \zeta \dot{d} \: : \:\: \alpha = 2(1-d)\psi^{+}_{cr} + 2\psi_{*}d > 0 \:\: \text{with} \: \: \psi^{+}_{cr} >0, \: \psi_{*}>0, 
\label{omegadis expression}
\end{equation}

\noindent where $\psi^{*}$, $\psi^{+}_{cr}$ are coefficients with units of energy per unit volume, and $\zeta>0$ is a constant kinetic modulus so that the dissipation inequality is satisfied when $\dot{d}>0$. The timescale for damage growth is determined by $\zeta$, with decreasing values of $\zeta$ leading towards rate-independent damage evolution. $\psi^{+}_{cr} + \psi_{*}$ represents a part of the energy dissipated per unit volume as $d$ grows from 0 to 1. The microforce balance in \eqref{microscopic traction condition final xi} along with \eqref{state relation 2nd PK stress and vector microstress} and \eqref{omegadis expression} gives the evolution equation for $d$ as

\begin{equation}
    \zeta \dot{d} = 2(1-d)\mathcal{H} -2\psi_{*}\Big(d- l^{2} \dfrac{\partial^{2} d}{\partial X_{i} \partial X_{i}}\Big),
\label{final evolution equation for d m}
\end{equation}

\noindent where $\zeta>0$ is a constant kinetic modulus. $\zeta$ determines the timescale of damage growth with decreasing values of $\zeta$ leading towards rate-independent damage evolution. $\mathcal{H}$ is a monotonically increasing history function expressed as
\begin{equation}
\begin{split}
    & \hspace{3.5 cm} \mathcal{H}(t) \overset{\text{def}}{=} \underset{s \in [0,t]}{\text{max}} \Big[\langle \psi^{+}_{\text{o}}(s) - \psi^{+}_{cr} \rangle\Big] ,
 \\& \langle \psi^{+}_{\text{o}}(s) - \psi^{+}_{cr} \rangle = \begin{cases}
\begin{aligned} 
			&0 \hspace{0.2in} \hspace{0.1in} &&\text{if}  \hspace{0.1in} \psi^{+}_{\text{o}}(s) - \psi^{+}_{cr} < 0,\\
			& \psi^{+}_{\text{o}}(s) - \psi^{+}_{cr}  && \text{if} \hspace{0.1in} \psi^{+}_{\text{o}}(s) - \psi^{+}_{cr} \geq 0 \: \text{and} \: \psi^{(B)}_{\text{o}} \ge \psi^{(B)}_{cr}.
\end{aligned}
	\end{cases} 
\end{split}
\label{monotonically increasing history function m}
\end{equation}

\noindent Damage initiates only when $\psi^{+}_{\text{o}}$ reaches the critical value $\psi^{+}_{cr}$ (the energetic contribution to $\mathit{W}_{cr}$) along with $\psi^{(B)}_{\text{o}} = \psi^{(B)}_{cr}$.\footnote{A macroscopic stretch measure, e.g., $\overline{\lambda}(t/0)$ by itself, cannot be used as a criterion for damage initiation. This can be understood by considering a loading rate much slower than the dynamic crosslink macroscopic relaxation timescale ($\tau_r$). In this case, the polymer network can easily reconfigure. Even when a specimen is subjected to large global deformation, the subchains, due to continuous network reconfigurations, will be stretched by an insignificant amount. For loading rates much faster than $\tau_r$, the polymer network will not have sufficient time to reconfigure, and the subchains will undergo a significant amount of stretching upon large specimen deformation. Hence, to capture the extent of subchain stretching, $\overline{\lambda}(t/0)$ will have to be further supplemented with a measure of loading rate, e.g., $\dot{\overline{\lambda}}(t/0)$. $\overline{\lambda}(t/0)_{cr}$ will then not be a unique value but dependent on $\dot{\overline{\lambda}}(t/0)$. In contrast, $\psi^{+}_{cr}$ and $\psi^{(B)}_{cr}$ are unique values.} Here, material strength is explicitly accounted for. 

$\psi^{+}_{cr} + \psi_{*}$ represents a part of the energy dissipated per unit referential volume as $d$ grows from 0 to 1. $\psi_{*}$ denotes additional referential energy density dissipated as $d$ increases from 0 to 1. In this paper, we have focused on fracture under an overall tensile loading. For soft polymers with tension-compression response asymmetry, different values for the parameter set $\{\psi^{+}_{cr}, \psi_{*}, l, \zeta\}$ can be used under compressive loading. This change of parameter values can be triggered using the field variable, which governs the tension-compression response asymmetry in the constitutive model.

The complete set of model parameters for deformation, damage initiation, and damage growth is

\begin{equation}
    \underbrace{\{k_{s}, k_{ns}, n_{s}, G, K, \nu_0, m, \alpha_p, \mathsf{S}_0, h, \overline{\lambda}_L\}}_{\text{deformation}} ,\:\: \underbrace{\{\psi^{+}_{cr}, \psi_{*}, l, \zeta\}}_{\text{damage}}.
    \label{Calibration parameters}
\end{equation}

For incompressible elastomers, the list of model parameters is reduced to 

\begin{equation}
    \underbrace{\{n_{s}, \overline{\lambda}_L\}}_{\text{deformation}} ,\:\: \underbrace{\{\psi^{+}_{cr}, \psi_{*}, l, \zeta\}}_{\text{damage}}.
    \label{Calibration parameters elastomers}
\end{equation}

\section{Numerical procedure}
\label{Numerical procedure}

\subsection{Finite element implementation of the damage model}

\begin{figure}[h!]
    \begin{center}
		\includegraphics[width=\linewidth]{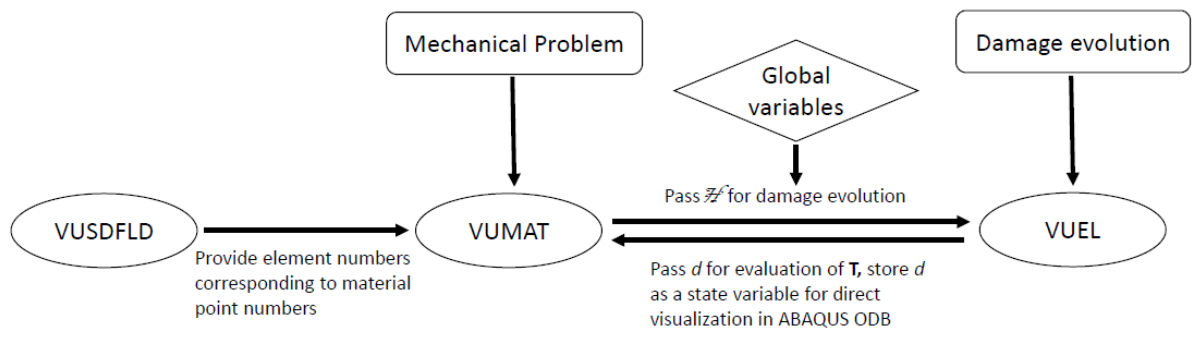}
	\end{center}
    \caption{{\textbf{Summary of the finite element implementation workflow.} The total Cauchy stress tensor $T_{ij}$ is denoted by $\textbf{T}$.}} 
    \label{Numerical implementation summary}
\end{figure}

The model was implemented in the finite element software ABAQUS, which has been acceptably used for engineering simulations involving complex loadings, inhomogeneous deformations, and non-linear material responses \citep{ABAQUSshabnamSEI, Niu2022SimulationMeasurements, ABAQUSvaishakh, Niu2022engwithcomp, Bai2021AEngineering,  Zhong2021AGrowth, Srivastava2011STRESSPIPE}.
The damage variable $d$ is an additional degree of freedom at each node along with displacement. Therefore, implementation in ABAQUS requires the development of user-defined elements. Since the timescales involved are short, an explicit formulation was used, and correspondingly, a VUEL (user-defined element with explicit formulation) was developed. Following the recent works outlining the numerical implementation of phase fields models using VUEL \citep{VUEL1, VUEL2, VUEL3}, we split the problem computationally into (i) a mechanical problem (VUMAT, user-defined mechanical material behavior with inbuilt elements) and (ii) a problem corresponding to the additional degree of freedom $d$ (VUEL). The evolution of $d$ at a node is dependent upon quantities evaluated in the mechanical problem (VUMAT), specifically the damage driving history function $\mathcal{H}$ at the corresponding material (integration) points. Further, to update the total Cauchy stress ${T}_{ij}$ at a material point, $d$ evaluated from the damage problem (VUEL) at the corresponding nodes is required. This information is passed between VUMAT and VUEL using global variables and the VUSDFLD subroutine. The finite element implementation workflow is summarized in Figure \ref{Numerical implementation summary}.

\subsection{Numerical update procedure for ${T}_{ij}^{(B)\text{o}}$ and $\psi^{(B)}_{\text{o}}$}
\label{Update procedure for cauchy stress and undamaged free energy}

The finite element program ABAQUS returns only the deformation gradients from the previous and current time steps at each material point. Direct numerical implementation of the analytical expressions for ${T}_{ij}^{(B)\text{o}}$ and $\psi^{(B)}_{\text{o}}$ in ABAQUS will require the storage of the entire deformation gradient history along with many state-dependent variables at each material point. The resulting storage requirement will be huge, making it challenging. A numerical update procedure for ${T}^{(B)\text{o}}_{ij}$ developed by Konale and co-workers \citep{KonalePBS} is applied, which requires deformation gradients only from the previous and current time steps at each material point. It is summarized as
\begin{equation}
\begin{split}
    & {T}^{(B)\text{o}}_{ij}(t+\Delta t) \rightarrow e^{-k_{ns} (t + \Delta t)} \mu {\Gamma}_{ij}(t+\Delta t/0) + {T}^{(B)\text{o}2}_{ij}(t+\Delta t), 
    \\& {T}_{ij}^{(B)\text{o}2}(t+\Delta t) \rightarrow e^{-k_{ns} \Delta t} J^{-5/3}(t+\Delta t/t) \: {F}_{ik}(t+\Delta t/t) \: {T}^{(B)\text{o}2}_{kl}(t) \: {F}_{jl}(t+\Delta t/t) \\ 
    & + \dfrac{1}{2} \dfrac{n_{ns}}{n_s} k_s e^{-k_{ns} \Delta t} \: \mu \: {\Gamma}_{ij}(t+\Delta t/t) \: \Delta t. 
\end{split}
\label{TB1 update procedure, final}
\end{equation}
We apply the procedure to $\psi_{\text{o}}^{(B)}$ which can be summarized as 
\begin{equation}
\begin{split}
     & \psi^{(B)}_{\text{o}}(t+\Delta t) \rightarrow e^{-k_{ns} (t+\Delta t)}  \mu \Bigr[\overline{\lambda}_L\Bigr]^2 \: \zeta(t+\Delta t/0) + \psi^{(B)}_{\text{o2}}(t+\Delta t) - \mu  \Bigr[\overline{\lambda}_L\Bigr]^2 \: \zeta(0), \\& \psi^{(B)}_{\text{o2}}(t+\Delta t) \rightarrow e^{-k_{ns}\Delta t} \psi^{(B)}_{\text{o2}}(t) + \dfrac{1}{2}\Big[\Big(\dfrac{n_{ns}}{n_s}  k_s e^{-k_{ns}\Delta t} \mu \Bigl[\overline{\lambda}_L\Bigr]^2 \: \zeta(t+\Delta t/t)\Big) \\& + \Big(\dfrac{n_{ns}}{n_s}  k_s \mu \Bigl[\overline{\lambda}_L\Bigr]^2 \: \zeta(t+\Delta t/t+\Delta t)\Big)\Big] \Delta t.
\end{split}         
\label{psiB1 update expression final}
\end{equation}

\subsection{Finite element model reduction using symmetries}

For all simulations in this paper, the bottom surface of the specimen is completely fixed, i.e., $u_{\text{X}}$ = $u_{\text{Y}}$ = $u_{\text{Z}}$ = $u^{\text{R}}_{\text{X}}$ = $u^{\text{R}}_{\text{Y}}$ = $u^{\text{R}}_{\text{Z}}$ = 0. $u$, $u_{\text{R}}$ denote the translational and rotational degrees of freedom of a material point about the axis indicated in the subscript. Z and Y axes are parallel to the thickness and height of the specimen, respectively. For quarter block simulations, $u_{\text{X}}$ = $u^{\text{R}}_{\text{Y}}$ = $u^{\text{R}}_{\text{Z}}$ = 0 is prescribed on the symmetry surface with normal along X axis. For both quarter and half block simulations, $u_{\text{Z}}$ = $u^{\text{R}}_{\text{X}}$ = $u^{\text{R}}_{\text{Y}}$ = 0 is prescribed on the symmetry surface with normal along the Z axis. Finally, for all simulations, $u_{\text{Y}}$ corresponding to the loading profile is prescribed along with $u_{\text{X}}$ = $u_{\text{Z}}$ = $u^{\text{R}}_{\text{X}}$ = $u^{\text{R}}_{\text{Y}}$ = $u^{\text{R}}_{\text{Z}}$ = 0 on the specimen's top surface.

\section{Model parameter calibration for PBS}
\label{Model Fit}

\begin{figure}[h!]
    \begin{center}
		\includegraphics[width=\textwidth]{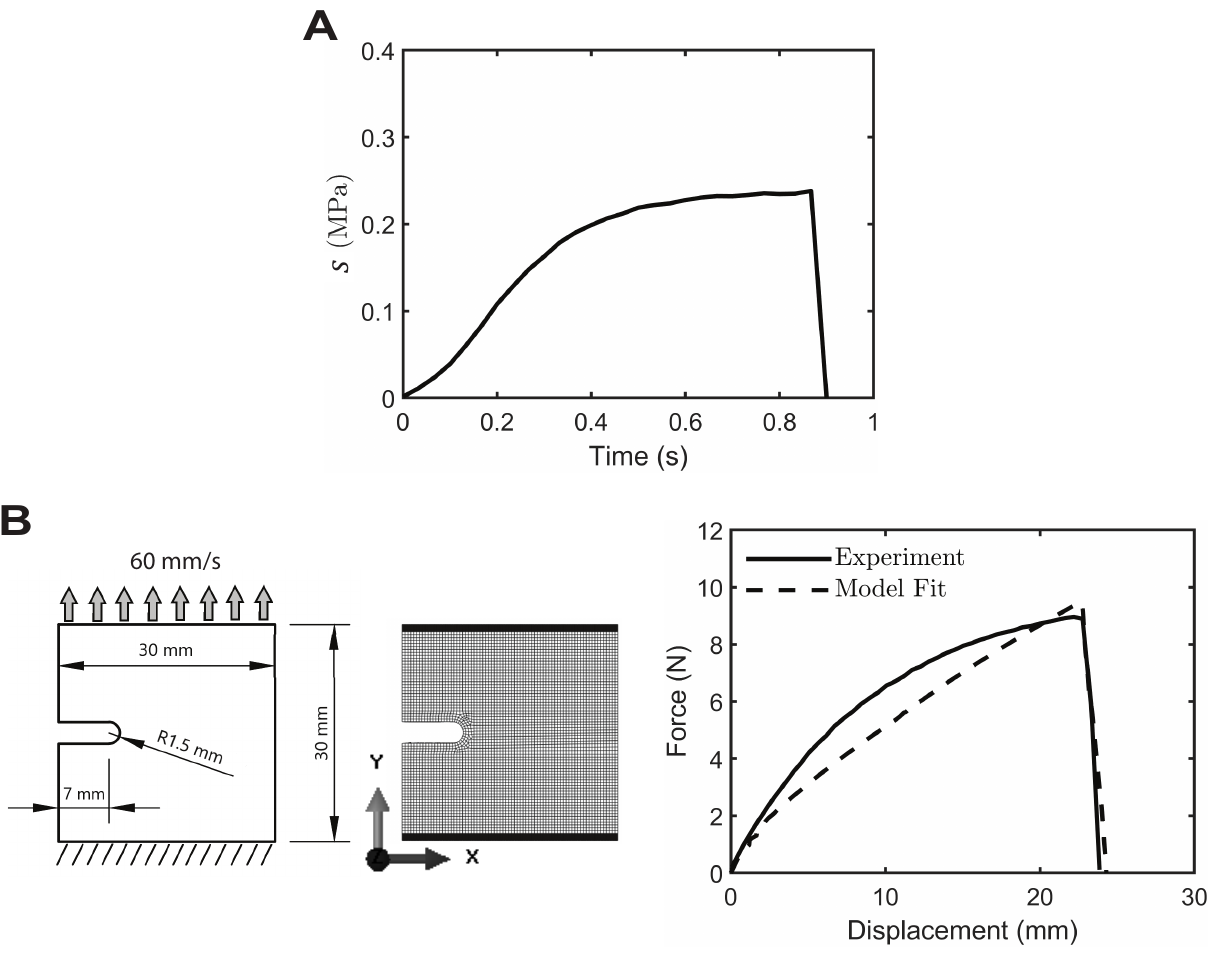}
	\end{center}
    \caption{{\textbf{Fracture model fit to experiments for PBS.} (\textbf{A}) The engineering stress ($s$) vs. time curve for the uniaxial tension experiment on PBS. The gauge section of the specimen undergoes stretching at an engineering strain rate of 4 s$^{-1}$. (\textbf{B}) The schematic for the notched PBS tension experiment. The specimen has a thickness of 3.2 mm and was subjected to a loading speed of 60 mm/s. The finite element mesh used for the simulation. The fit of the model (dashed lines) to the experimental extension force-displacement curve.}}
    \label{Calibration}
\end{figure}

Values for all the deformation-associated parameters in \eqref{Calibration parameters} are taken from \citep{KonalePBS}. Uniaxial homogeneous tension tests can be used for evaluation of $\mathit{W}_{cr}$, $\psi^{+}_{cr}$ and $\psi^{(B)}_{cr}$.\footnote{Soft polymers exhibit fast damage evolution through sharp load drops corresponding to fracture \citep{ZehnderkineticfracturePVA, fastdamagevolution1}. Hence, $\mathit{W}_{cr}$ (note that it accounts for both energetic and dissipative contributions) for soft polymers will be approximately equal to the value of $\mathit{W}$ at damage initiation ($\psi^{+}_{\text{o}}$ = $\psi^{+}_{cr}$). For hyperelastic materials, $\mathit{W}_{cr}=\psi^{+}_{cr}$ (the energetic contribution) since there are no viscous contributions to $\mathit{W}_{cr}$. $\mathit{W}_{cr}$ $>$ $\psi^{+}_{cr}$ for materials with viscous effects.} Dogbone tensile PBS specimens of ASTM D638 Type V standard [gauge dimensions: 9.5$\pm$1 mm x 3.2$\pm$1 mm x 3.2$\pm$1 mm] \citep{ASTMD638} were subjected to a stretching speed of 60 mm/s. Displacement was monitored through markers at the two extremes of the gauge length and standard image processing. \eqref{Stress work definition} for this loading can be written as 

\begin{equation}
    \mathit{W}({X}_{i}, t) = \int_{0}^{t} s \dot{\lambda} dt
\label{W_cr uniaxial tension}
\end{equation}

\noindent where $\mathit{s}$, $\lambda$ are the engineering stress and gauge section stretch in the direction of loading, respectively.\footnote{Analytical expressions for $\mathit{W}_{cr}$ in the case of uniaxial tension can be obtained for hyperelastic materials using \eqref{W_cr uniaxial tension}. E.g., if neo-Hookean model is used: $\mathit{W}_{cr} = \mu(\frac{\lambda_{cr}^2}{2} + \frac{1}{\lambda_{cr}} - \frac{3}{2})$, if Mooney-Rivlin model is used: $\mathit{W}_{cr} = 2 C_{1}(\frac{\lambda_{cr}^2}{2} + \frac{1}{\lambda_{cr}} - \frac{3}{2}) + 2 C_{2} (\lambda_{cr} + \frac{1}{2\lambda^{2}_{cr}} - \frac{3}{2})$ where $C_{1}$, $C_{2}$ are material constants and the ground state shear modulus $\mu = 2(C_{1} + C_{2})$.} $\dot{\lambda}$ was evaluated to be 4 s$^{-1}$ and the corresponding $\sigma$ vs. $t$ curve is shown in Figure \ref{Calibration}A. $\mathit{W}_{cr}$ = 0.6 MPa was calculated using \eqref{W_cr uniaxial tension} with the integration carried till the time instant of complete failure. A one-dimensional version of the model without considering damage was implemented in MATLAB to simulate this experiment and calculate $\psi^{+}_{\text{o}}$ and $\psi^{(B)}_{\text{o}}$. $\psi^{+}_{cr}$ = 0.27 MPa, $\psi^{(B)}_{cr}$ = 0.12 MPa were obtained as the values of $\psi^{+}_\text{o}$, $\psi^{(B)}_\text{o}$ at complete failure. 

Although the parameters $\psi_{*}, l, \zeta$ can be fit to the uniaxial tensile test, sensitivity to damage gradients can only be obtained with inhomogeneous deformations, e.g., the notched tension test in Figure \ref{Calibration}B. The finite element mesh with a total of 32,468 C3D8R (three-dimensional, reduced integration) stress elements used for simulating the notched tension experiment is shown in Figure \ref{Calibration}B. In all the simulations in this work, a finer mesh was applied in the regions surrounding the notch to capture the stress and strain gradients. Several works in literature have studied failure in soft polymers experimentally \citep{Npm2units1, Macroscopicfractureprocesszone2} and theoretically \citep{Macroscopicfractureprocesszoneth1, Macroscopicfractureprocesszoneth2} by considering a macroscopic fracture process zone. The $l$ value based on the length scale of the fracture process zone in a gradient-damage-based model for elastomers (similar to the gradient-damage model in this work but limited to hyperelastic materials) has been shown to predict size-dependent fracture \citep{Parabolicshape2}. The actual value of the length scale associated with the macroscopic fracture process zone in PBS has not been experimentally measured so far. Hence, the parameter $l$ is considered a regularization parameter here. $l$ = 2.5 mm was chosen to result in a computationally tractable mesh given the specimen dimensions while following the restriction of $l \underset{\approx}{>} 5 \:\: \text{x} \:\: \text{typical element size} \: (h_{e})$ \citep{Anand(2021)flowrule} and being in the range of fracture process zone length scales reported for soft polymers in literature \citep{Damagelengthscale, Fracturechallenges2, Damagelengthscale2}.  $\psi_{*}$ and $\zeta$ were calibrated to fit the load-displacement curve after the load maximum in the notched tension calibration experiment. Model fit to the experimental extension force-displacement curve corresponding to the notched tension test is shown in Figure \ref{Calibration}B. Values of the model parameters are listed in Table 1. For PBS, we assume $k_{ns}=k_{s}=k$.

\begin{table}[!h]
\begin{center}
\begin{tabular}{ll}
\begin{tabular}{ | m{2.5cm} | m{2 cm} | } 

  \hline

  \multicolumn{2}{|c|}{Deformation} \\
  
  \hline
  Parameter & PBS\\

  \hline

  $k$ (s$^{-1}$) & 0.35 \\

  $n_s$ (m$^{-3}$) &  6.4 x 10$^{24}$\\
  
  \hline

   $G$ (MPa) & 0.4\\
   
   $K$ (MPa) & 20\\ 
  
   \hline

  $\nu_0$ (s$^{-1}$) & 2 x 10$^{-3}$\\
   
  $m$ (-) & 0.95\\

  $\alpha_p$ (-) & 0.11\\

  $\mathsf{S}_0$ (kPa) & 0.6 \\
  
  $h$ (kPa) & 37.7\\
  
  \hline

  \vspace{0.1cm}
 
  $\overline{\lambda}_L$ (-) & 37.4\\

  \hline
\end{tabular}
&
\begin{tabular}{ | m{2.5cm} | m{2 cm} | } 

  \hline

  \multicolumn{2}{|c|}{Damage} \\

  \hline
  
  Parameter & PBS\\

  \hline

  $\psi^{+}_{cr}$ (MPa) &  0.27\\

  $\psi_{*}$  (kPa) &   4\\

  $l$ (mm) &   2.5\\

  $\zeta$ (kPa.s) &   1.1\\

  \hline
\end{tabular}
\end{tabular}
\caption{\textbf{Model parameters for PBS.}}
\end{center}
\label{Material parameters} 
\end{table}

\section{Model predictions for fracture in a variety of soft polymers}

The model is applied to predict fracture in select soft polymers - a permanently crosslinked rubber [ethylene propylene diene monomer (EPDM)], PBS, and dynamically crosslinked EPS25 vitrimer. Also, we show that macroscopic model parameters for PBS can estimate subchain dissociation energy, which is a fracture microstructural process-related quantity. Three independent experiments were performed on PBS with different specimen geometries and loading configurations subjected to a range of stretching speeds. The previously calibrated model parameters for PBS obtained using the notched tension experiment in Figure \ref{Calibration}B were kept fixed for independent validations. This allowed for an independent, true assessment of the model's predictive capabilities for the three validation experiments on PBS. Experimental data reported in \citep{EPDMrubber} and \citep{Shenmodeling} is used to test the predictive abilities of the proposed model for EPDM rubber and EPS25 vitrimer, respectively. Based on the experimental material response, intermolecular resistance (mechanism $\mathit{A}$) is neglected for modeling the deformation of EPDM rubber and EPS25 vitrimer in \citep{EPDMrubber, Shenmodeling}, i.e., $\psi_\text{o}^{(A)}=0$ and $\psi_{\text{o}} = \psi_\text{o}^{(B)}$. For the network resistance (mechanism $\mathit{B}$), the free energy density functions with the dynamic crosslink formulation for the hyperelastic function part were the same as used in \citep{EPDMrubber, Shenmodeling} for EPDM rubber, EPS25 vitrimer. This also allowed us to test the dynamic crosslink formulation's predictive capabilities for different hyperelastic free energy density base functions. Deformation-related material parameters for EPDM and EPS25 vitrimer are taken from \citep{EPDMrubber, Shenmodeling}.

\subsection{PBS: Estimate of subchain dissociation energy using $\psi^{+}_{cr}$}

\begin{figure}[h!]
    \begin{center}
		\includegraphics[width=\textwidth]{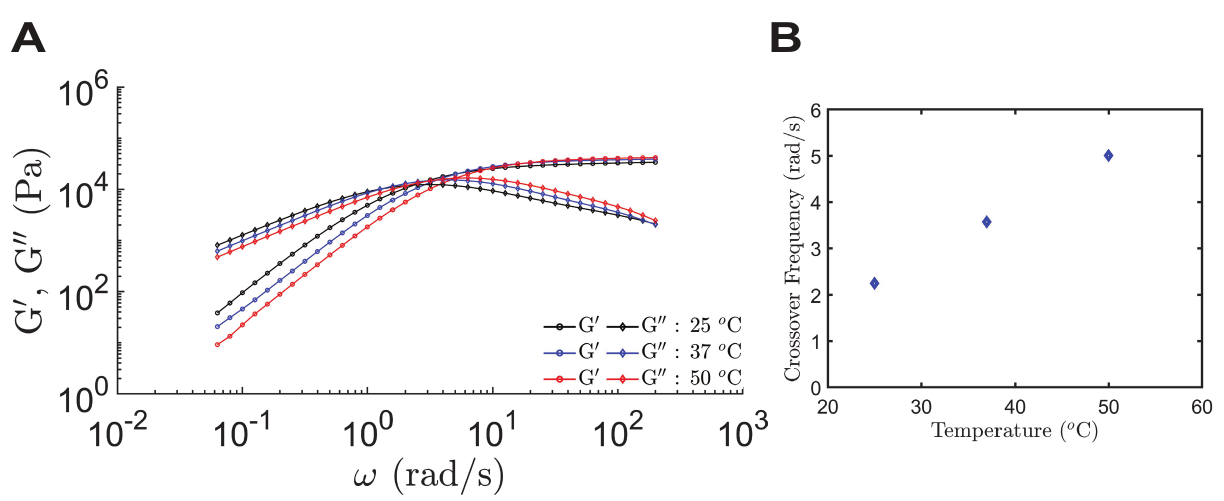}
	\end{center}
    \caption{{\textbf{Oscillatory rheology experiments on PBS to validate subchain dissociation energy estimate obtained using $\psi^{+}_{cr}$.} (\textbf{A}) The oscillatory shear response of PBS (1\% shear strain) at 25 $^{\text{o}}$C, 37 $^{\text{o}}$C and 50 $^{\text{o}}$C. (\textbf{B}) The crossover frequency vs. temperature plot for PBS.}}
    \label{PBSRheology}
\end{figure}

The referential number density of subchains in PBS surviving from $t$=0 till any time instant $t$ \Big($n_{s}(t/0)$\Big) can be obtained using the first order kinetics of underlying dynamic reactions \citep{KonalePBS} as equal to $n_{s} e^{-k_{ns}t}$. At the time instant corresponding to the load drop in Figure \ref{Calibration}A for the uniaxial tension test, with $k_{ns}=0.35$ s$^{-1}$ and $n_s$ = 6.4 x 10$^{24}$ m$^{-3}$, $n_{s}(t/0)$ is equal to 4.7 x 10$^{24}$ m$^{-3}$. These subchains experience the highest level of stretch and, hence, will be most likely to undergo stretch-induced dissociation. Hence, it can be assumed that the energy associated with the dissociation of 4.7 x 10$^{24}$ m$^{-3}$ subchains at a material point is $\psi^{(B)}_{cr}$. The subchain dissociation energy can then be estimated as 15 kJ/mol. We independently conducted oscillatory rheology experiments on PBS at temperatures of 25 $^{\text{o}}$C, 37 $^{\text{o}}$C and 50 $^{\text{o}}$C shown in Figure \ref{PBSRheology}A till 1 \% strain. The experiments were performed using the ARES G2 Rheometer and the cone and plate geometry (40 mm diameter and 0.04 radians cone angle). Three repetitions were performed for each experimental result reported. Konale and co-workers \citep{KonalePBS} established that the crossover frequency in small-strain frequency sweep experiments is proportional to the backward reaction rate parameter ($k_{ns}$) for the reversible reaction involving subchain dissociation in \eqref{generalized reversible crosslink reaction}. Assuming Arrhenius temperature dependence of $k_{ns}$, the activation energy for the backward reaction, i.e., subchain dissociation, can be obtained by fitting to the crossover frequencies in Figure \ref{PBSRheology}B as 24 kJ/mol, which is in the same order of magnitude as the subchain dissociation energy (15 kJ/mol). This same order of magnitude similarity between activation energy and subchain dissociation energy is reasonable considering (a) the relative simplicity of the chemical reaction under consideration - only one type of bond (B:O coordinate-bond) is involved, and (b) the assumptions made during the estimation analyses. Further, the subchain dissociation energy value lies in the range of weak, non-covalent bond energies reported in the literature \citep{Noncovalentenergyref1, Noncovalentenergyref2, Noncovalentenergyref3}. A physical understanding of the macroscopic fracture model parameters can thus provide reasonable approximations for fracture-related microstructural processes.

\subsection{PBS: Single edge notch test for validation}
\label{18.5 mm x 3 mm insert one side}

\begin{figure}[h!]
    \begin{center}
        \includegraphics[width=0.73\textwidth]
        {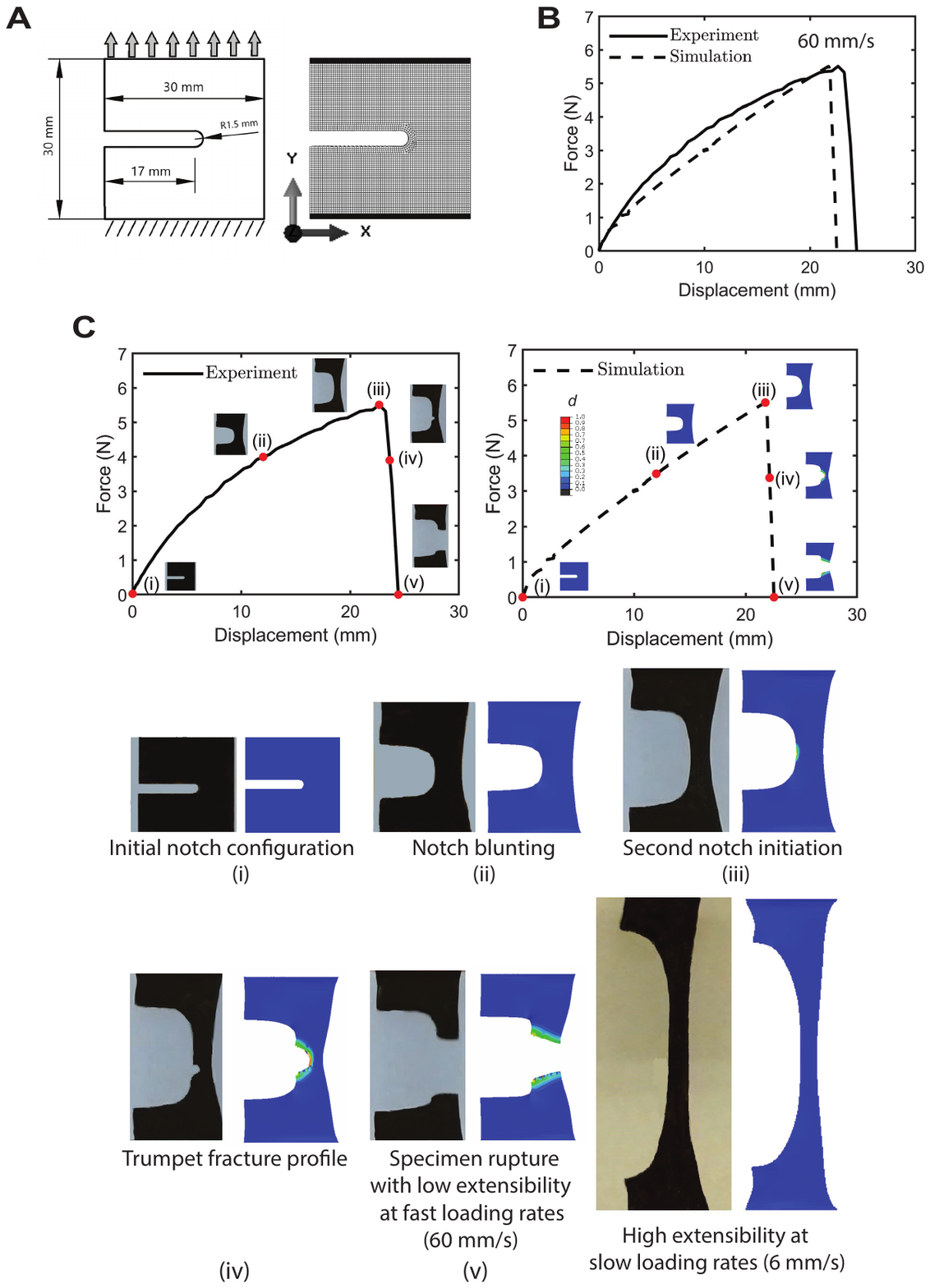}
	\end{center}
    \caption{{\textbf{Predictive ability of the fracture model for PBS (forces and rate-dependent physical response).} (\textbf{A}) The edge notch PBS stretch experiment schematic. The finite element mesh used for the simulations. (\textbf{B}) A comparison between the numerically predicted force-displacement curve and the experimental result for 60 mm/s loading speed shows good agreement. (\textbf{C}) The specimen shapes at different levels of stretch from the experiment (left) and the simulation (right) show good agreement for the 60 mm/s stretching speed. The model predicts the experimentally observed trumpet fracture profile. Also, the model predicts rupture with low extensibility at a fast loading rate (60 mm/s) and high extensibility at a slow loading rate (6 mm/s)}. Elements with $d>0.95$ are hidden, and the experimental specimen was colored for visualization.} 
    \label{Validation 1}
\end{figure}

A U-shaped edge notch with depth and radius of 18.5 mm and 3 mm, respectively, was created equidistant from the gripping fixtures in the gauge section with dimensions 30$\pm$2 mm x 30$\pm$2 mm x 3.2$\pm$1 mm. This is a much deeper notch than what was used for calibration. The specimen geometry for the experiment schematic shown in Figure \ref{Validation 1}A was stretched at 60 mm/s. The finite element mesh (31,892 C3D8R elements) used for simulation is shown in Figure \ref{Validation 1}A. The specimen was stretched until failure. Good agreement between the prediction from simulation and experimental result for the extension force can be seen in Figure \ref{Validation 1}B. Specimen geometry changes at different displacement levels from the experiment and simulation are shown in Figure \ref{Validation 1}C, which also shows very good agreement between the experimental and simulation results. Particularly, the trumpet fracture profile (formation of second local notch ahead of the original blunted notch) observed experimentally in Figure \ref{Validation 1}C[iv] is predicted by the model. The model correctly predicts the two cases of rupture with low extensibility at a fast loading rate (60 mm/s) and high extensibility at a slow loading rate (6 mm/s), as shown in Figure \ref{Validation 1}C. The high extensibility at the relatively slower loading speed of 6 mm/s is due to $\psi^{+}_\text{o}$, $\psi_\text{o}^{(B)}$ at the notch tip not being able to exceed $\psi^{+}_{cr}$ and $\psi^{(B)}_{cr}$, respectively.

\subsection{PBS: Center notch for validation}

\label{10 mm x 3 mm slot at 2 mm/s and 6 mm/s}

\begin{figure}[h!]
    \begin{center}
		\includegraphics[width=0.67\textwidth]{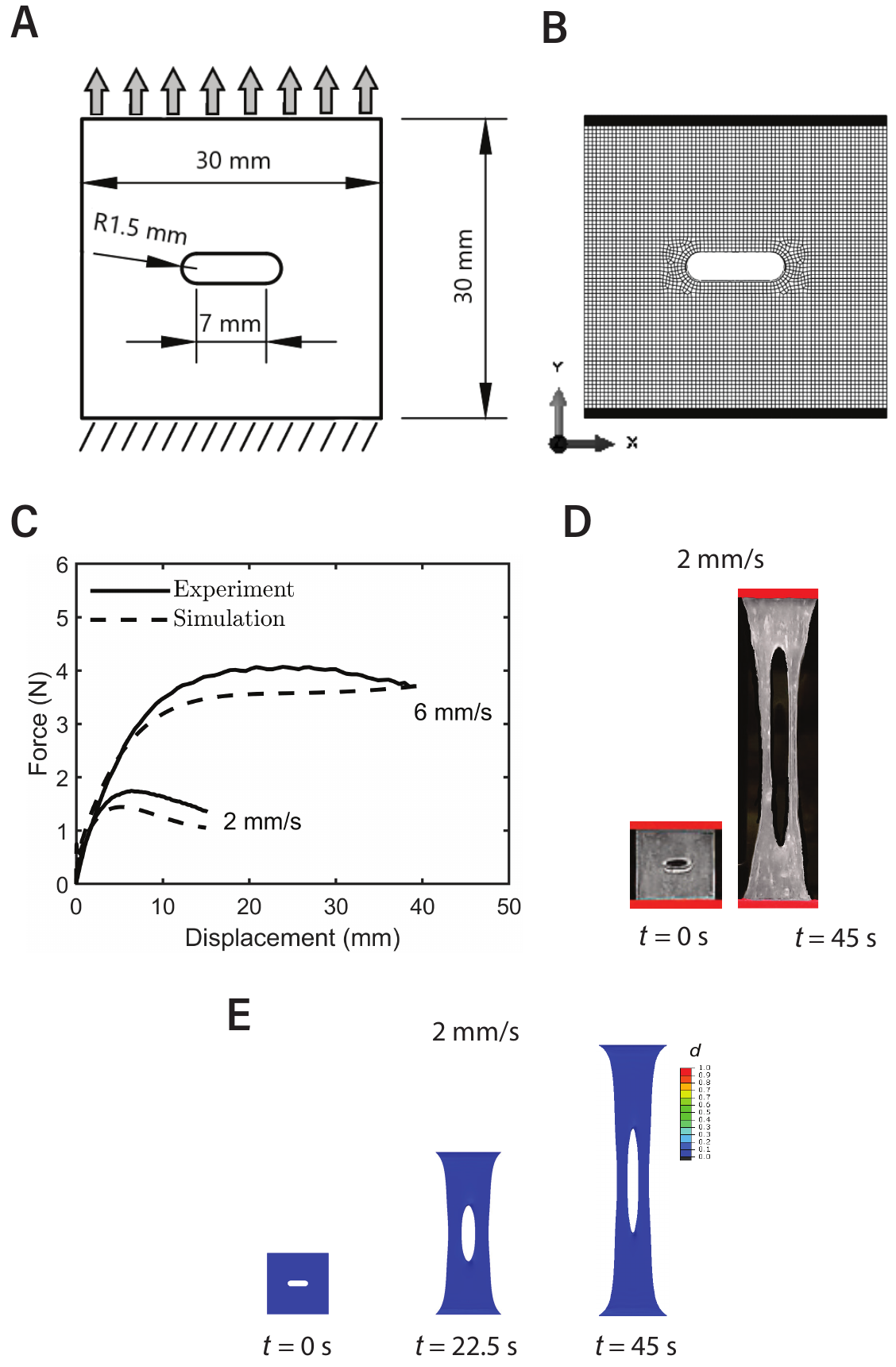}
	\end{center}
    \caption{{\textbf{Model predictions for center notch PBS specimen.} (\textbf{A}) The center notch PBS stretch experiment schematic. (\textbf{B}) The finite element mesh for the simulations. Quarter symmetry with appropriate symmetry boundary conditions was used in the simulations, but for clarity, the complete geometry with mesh is shown. (\textbf{C}) A good agreement between the model predicted and experimentally measured forces can be seen for the two loading speeds of 2 mm/s and 6 mm/s. Necking instability seen in the experimental results is predicted by the model. Note that fracture does not occur at these slower loading rates. (\textbf{D}) The high extensibility is shown by comparing experimental specimen shapes at $t$ = 0 s and $t$ = 45 s. (\textbf{E}) The deformed shape predictions from the simulation for the 2 mm/s speed at different stretch levels.}} 
    \label{Validation 2}
\end{figure}

To validate the model predictions for PBS in a scenario involving a different notch geometry (same gauge section dimensions and boundary conditions as the single edge notch geometry), a 10 mm x 3 mm slot was created at the center of the gauge section, as shown in Figure \ref{Validation 2}A. Further, relatively low loading speeds of 2 mm/s and 6 mm/s were used for this geometry. The finite element mesh used for the simulations is shown in Figure \ref{Validation 2}B. One-quarter of the specimen (25,832 C3D8R elements) was simulated with appropriate boundary conditions considering the symmetries of the problem. For these relatively slower speeds, the experimental results exhibit necking instability in the form of the non-positive slope of force-displacement curves after a certain level of displacement. This can be attributed to significant network reconfiguration occurring over these long timescales (relative to the B:O coordinate bond dynamic crosslink macroscopic relaxation timescale) due to dynamic crosslink relaxation and reformation. Model predictions for the force-displacement curves, particularly the displacement levels for the onset of necking instability, agree well with the experimental results as seen in Figure \ref{Validation 2}C. Fracture does not occur for these slow loading speeds which can be attributed to $\psi^{+}_\text{o}$ and $\psi_\text{o}^{(B)}$, not being able to exceed $\psi^{+}_{cr}$ and $\psi^{(B)}_{cr}$, respectively at the notch tip. The specimens exhibited high extensibility in the experiments, as shown in Figure \ref{Validation 2}D by comparing the undeformed and deformed (at $t=45$ s) shapes from the experiment for the 2 mm/s loading speed. Figure \ref{Validation 2}E shows the simulation-predicted deformed shapes at different levels of stretch corresponding to the stretching speed of 2 mm/s.

\subsection{PBS: Stretching using inserts for validation}

\begin{figure}[hbt!]
    \begin{center}
		\includegraphics[width=0.85\textwidth]{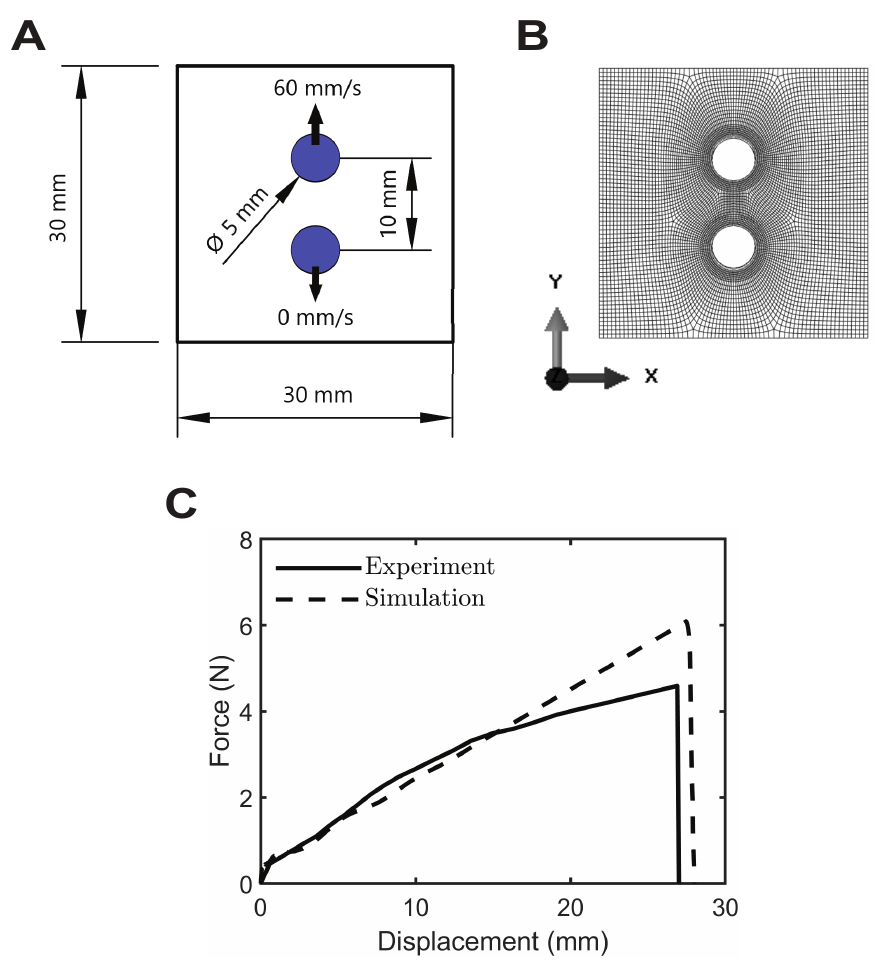}
	\end{center}
    \caption{{\textbf{Model prediction for stretching through internal holes in PBS specimen.} (\textbf{A}) The PBS specimen geometry and experiment schematic. The blue regions denote the inserts. The specimen has a thickness of 3.2 mm, and the top insert was displaced at a speed of 60 mm/s. (\textbf{B}) The finite element mesh used for the simulation. A quarter of the block is considered due to the symmetries of the problem with appropriate boundary conditions. For clarity, complete geometry with mesh is shown. (\textbf{C}) The model prediction for the extension force agrees well with the experimental result.}} 
    \label{Validation 3}
\end{figure}

The fracture model is now applied to a different loading configuration. The PBS specimen has two 5 mm diameter symmetric holes in the gauge section (same dimensions as the single edge notch geometry) as shown in Figure \ref{Validation 3}A. 3D-printed hard plastic (PLA) inserts of the same diameter placed in the holes are then used to stretch the specimen. The bottom insert is held fixed while the top insert is displaced at 60 mm/s. Figure \ref{Validation 3}A shows the experiment schematic. The finite element mesh used for the simulation is shown in Figure \ref{Validation 3}B. One-quarter of the specimen (29,016 C3D8R elements) was simulated with appropriate boundary conditions considering the symmetries of the problem. The specimen was stretched until failure.  The inserts were modeled as rigid analytical surfaces. Contact between the inserts and the specimen was assumed to be frictionless as the interfaces were lubricated in the experiments. Model prediction for the extension force agrees reasonably with the experimental result, as shown in Figure \ref{Validation 3}C. 

\subsection{Fracture response considering viscous effects}

\begin{figure}[hbt!]
    \begin{center}
		\includegraphics[width=0.8\linewidth]{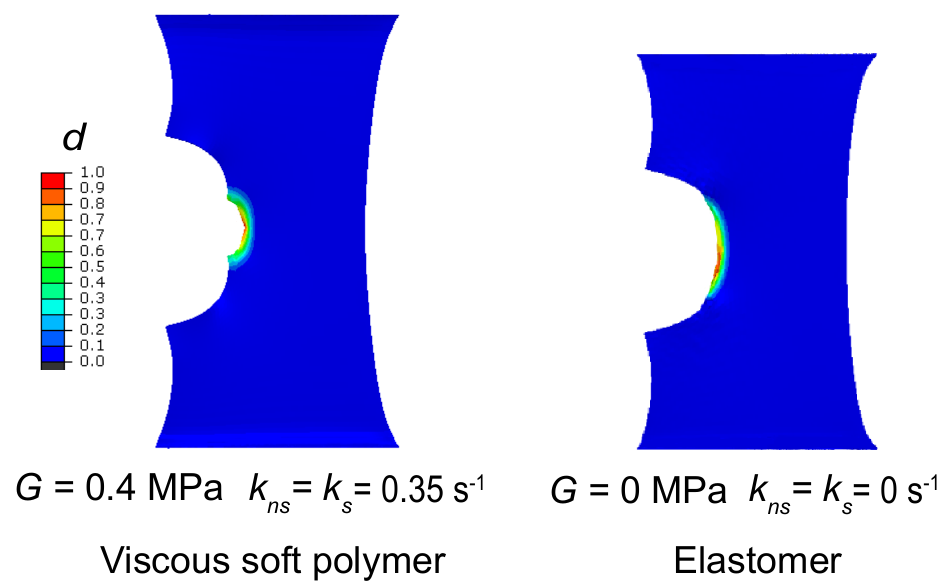}
	\end{center}
    \caption{{\textbf{Comparison of viscous soft polymer and elastomer fracture.} The notch profiles for the single edge notch specimen (Figure \ref{Calibration}B) stretching simulations showing experimentally observed trumpet profile for a viscous soft polymer and a single parabola-like profile when viscous effects were removed in the model to simulate the elastomeric material response. The profiles are shown just before complete failure. Elements with $d>0.95$ are hidden for visualization.}} 
    \label{Fracture shapes}
\end{figure}

We performed simulations where the viscous effects in the input material were removed (by suitably prescribing the material parameter values) to study the role of viscosity in the physical fracture phenomenon. These enabled qualitative model predictions for fracture profiles as the material varies from a viscous soft polymer to an elastomer. The geometry, loading schematic, and finite element mesh for the notched tension test shown in Figure \ref{Calibration}B were considered with a stretching speed of 60 mm/s. The material parameters for PBS were used for the viscous soft polymer (except $\zeta$) and elastomer (except $G, \: k_{ns}$, $k_{s}$, and $\zeta$).  We removed viscous effects by setting $G$ = 0 (removing the elastic-viscoplastic $A$ micro-mechanism) and $k_{ns} = k_{s}$ = 0 (this provides a model representing permanently crosslinked material) to simulate a hyperelastic material. The parameter $\zeta$, which determines the timescale for damage growth, was set equal to 3 kPa.s for both viscous soft polymer and elastomer for ease in elastomer numerical calculations. This numerical example shows differences in the general elastomer and viscous soft polymer fracture response. The computationally obtained fracture profiles in Figure \ref{Fracture shapes} transition correctly from the experimentally observed trumpet profile in viscous soft polymers \citep{Viscoelastictrumpet2, Shenmodeling} to the single parabola-like profile in elastomers \citep{Parabolaelastic1, Parabolaelastic2}.

The strains, and hence, the strain rates, are inhomogeneous around the notch tip. The simulations show that when there is local damage at the notch tip, the strains are further localized at the tip, and the strain rate increases significantly. In the case of rate-dependent viscous polymers, significantly higher strain rates around the notch tip increase the local stress, causing a small area near the notch tip to be damaged. This leads to the formation of a second, smaller notch and a trumpet-like profile that we observe in the experiments. Similar discussions about the trumpet fracture profile exist in the literature \citep{Viscoelastictrumpet1, Trumpetexpl1, Trumpetexpl2, Trumpetexpl3}. On the other hand, the elastomers are rate-independent, and there is no additional rate-dependent increase in stresses at the notch tip. As a result, the fracture progresses in a self-similar manner, i.e., the single parabola-like profile in elastomers. The model can predict the significantly varying important physical phenomena observed during the fracture of soft polymers.

\subsection{EPDM elastomer with V notch}

\begin{figure}[h!]
    \begin{center}
		\includegraphics[width=0.95\textwidth]{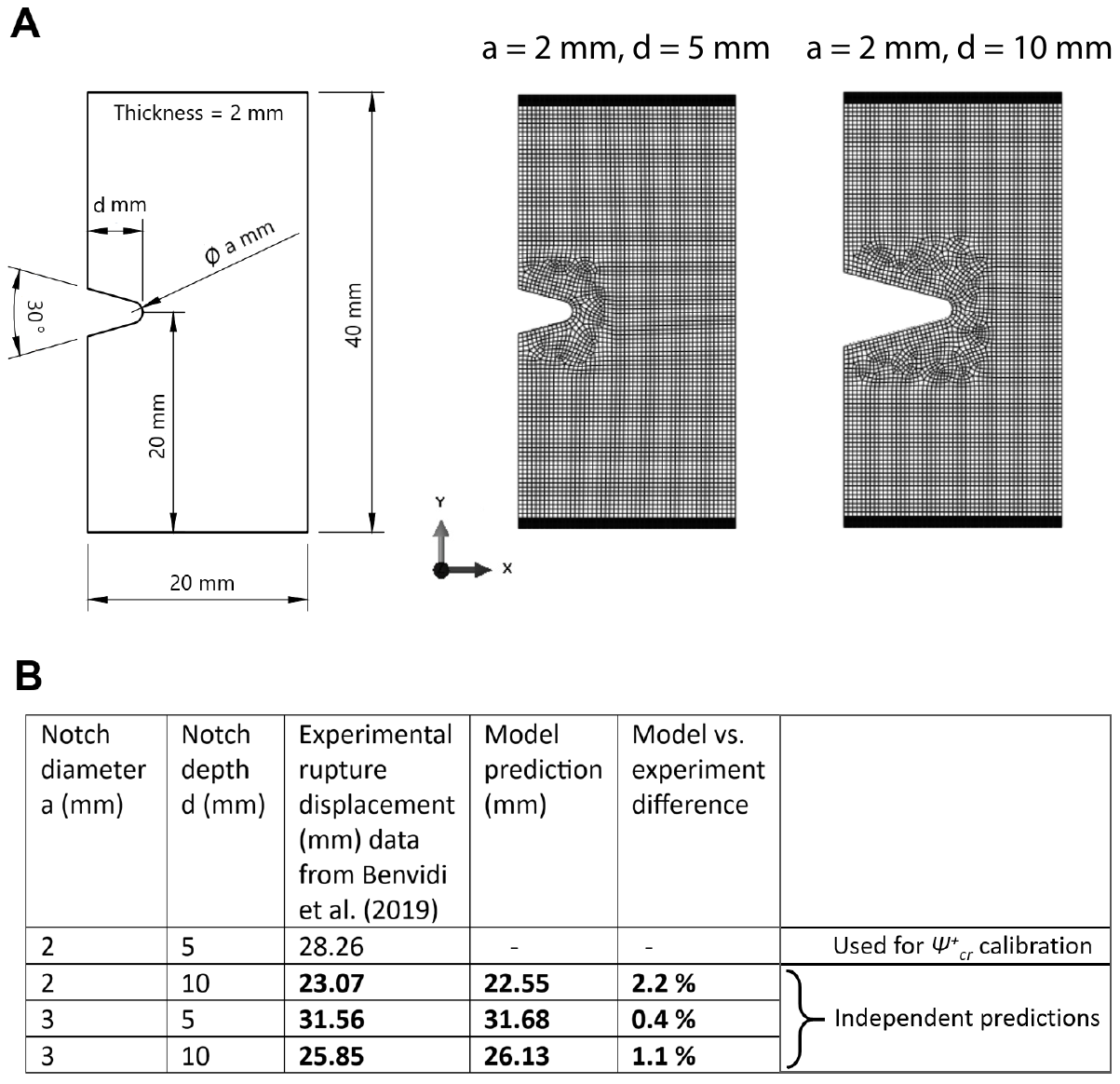}
	\end{center}
    \caption{{\textbf{Model prediction for fracture of EPDM elastomer.} (\textbf{A}) The V-notched specimen geometry for EPDM rubber used in \citep{EPDMrubber}. The finite element meshes used for the simulations of the a=2 mm d=5 mm, a=2 mm d=10 mm notch geometries.} (\textbf{B}) The experimental data for a=2 mm, d=5 mm from \citep{EPDMrubber} was used for $\psi^{+}_{cr} = \psi^{(B)}_{cr} = \mathit{W}_{cr}$ calibration. Independent model predictions for the rupture displacements for a=2 mm d=10 mm, a=3 mm d=5 mm and a=3 mm d=10 mm notch geometries show good agreement with the experimental results in \citep{EPDMrubber}.} 
    \label{EPDMvalidation}
\end{figure}

Benvidi and co-workers \citep{EPDMrubber} performed tension tests on V-notched EPDM rubber to demonstrate the efficacy of the average strain energy density criterion for fracture analysis of hyperelastic materials. We test our fracture model's ability to capture the experimental data reported in \citep{EPDMrubber}. The four notch configurations from \citep{EPDMrubber} were applied in finite element simulations for our model's verification. These configurations encompass variations in the depth and radius of the notch. One case (notch diameter a=2 mm, notch depth d=5 mm) was used for model parameter calibration. Fixing the values of these parameters, the remaining three cases were used for independent model validation. The finite element meshes for a=2 mm, d=5 mm and a=2 mm, d=10 mm cases are shown in Figure \ref{EPDMvalidation}A. A total of 20265, 20019, 20241, and 19746 C3D8R elements were used in the finite element meshes for a=2 mm d=5 mm, a=2 mm d=10 mm, a=3 mm d=5 mm, and a=3 mm d=10 mm cases, respectively. 

Due to the assumed hyperelastic nature of EPDM in \citep{EPDMrubber}, $\mathit{W}_{cr} = \psi^{+}_{cr} = \psi^{(B)}_{cr}$. The Ogden free energy density function was used in \citep{EPDMrubber} to model the large deformation response of the EPDM elastomer. Instead of the eight-chain free energy density function with the dynamic crosslink formulation for Mechanism $B$, the Ogden function is used with the same dynamic crosslink formulation for our fracture modeling framework. We set $G$ = 0 MPa, $k_{ns}$ = $k_{s}$ = 0 s$^{-1}$ in our model and used the deformation-related material parameters from \citep{EPDMrubber}. As the force-displacement curves for the V notched tests are not presented in \citep{EPDMrubber}, it was assumed that rupture occurs instantly when $\psi^{+}_{\text{o}} = \psi^{(B)}_{\text{o}}= \psi^{+}_{cr}$ at the notch tip. Independent predictions for the displacement at which rupture occurs for the a=2 mm d=10 mm, a=3 mm d=5 mm, and a=3 mm d=10 mm notch geometries (variations in notch depth and radius) agree well with the experimental results as shown in Figure \ref{EPDMvalidation}B.

\subsection{EPS25 vitrimer: Notched pure shear}

\begin{figure}[h!]
    \begin{center}
		\includegraphics[width=0.81\textwidth]{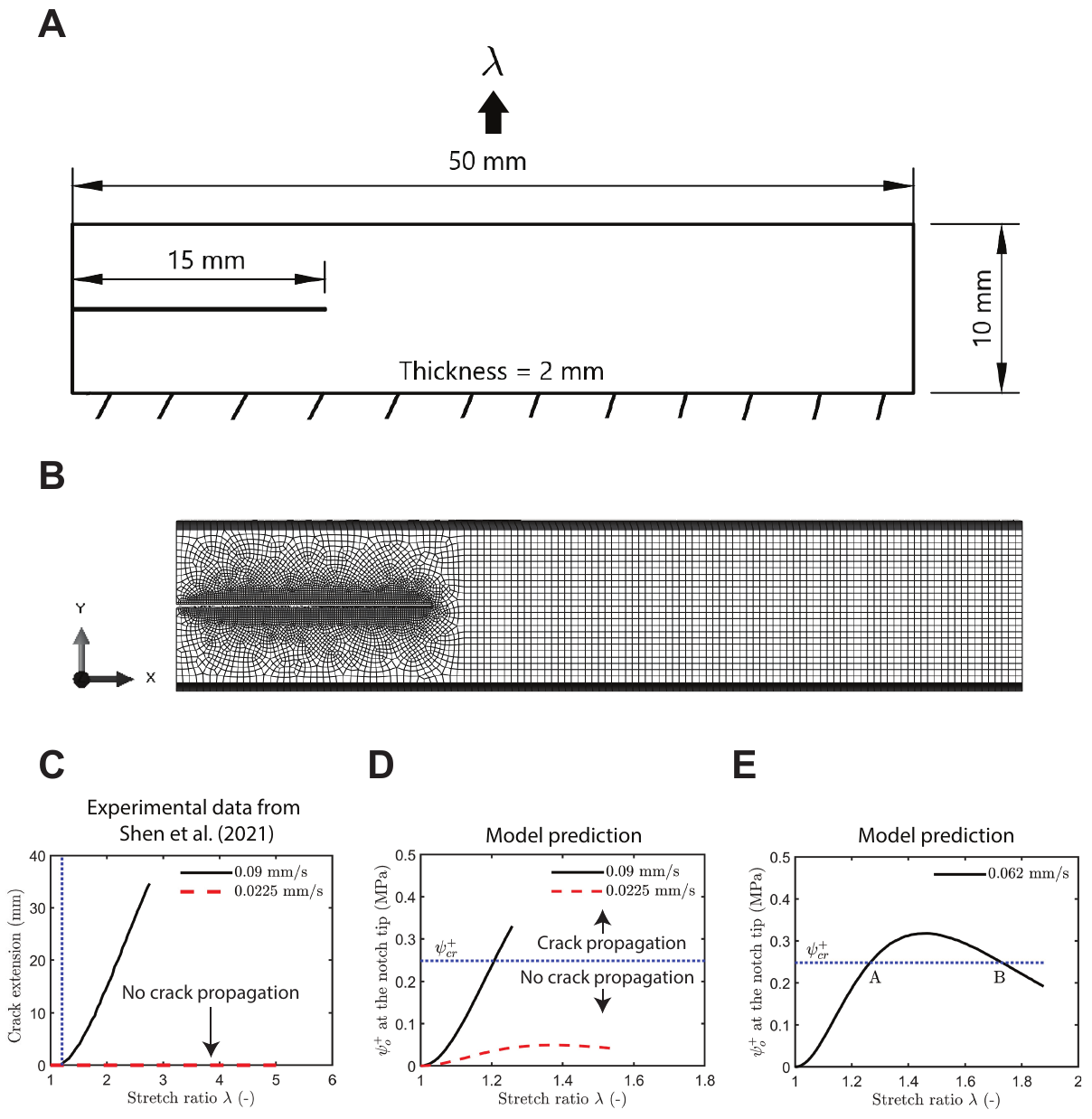}
	\end{center}
    \caption{{\textbf{Model prediction for fracture of EPS25 viscous soft polymer.} (\textbf{A}) The specimen geometry for EPS25 vitrimer pure shear tests with sharp single edge notches from \citep{Shenmodeling}. (\textbf{B}) The finite element mesh used for the simulations. (\textbf{C}) The crack extension vs. stretch ratio ($\lambda$) experimental data from \citep{Shenmodeling} for the fast and slow loading speeds of 0.09 mm/s and 0.0225 mm/s. There is no crack propagation for the 0.0225 mm/s loading speed. $\psi^{+}_{cr}=\psi^{(B)}_{cr}$ and $\mathit{W}_{cr}$ were calibrated using the $\lambda$ at initiation of crack propagation for 0.09 mm/s speed. (\textbf{D}) The model predictions for $\psi^{+}_{\text{o}}$ at the notch tip vs. $\lambda$ curves for the fast and slow loading speeds. The model predicts no crack propagation for the 0.0225 mm/s speed due to $\psi^{+}_{\text{o}}$ at the notch tip not being able to exceed $\psi^{+}_{cr}$. This agrees with the experimental observation in \citep{Shenmodeling}. (\textbf{E}) The model prediction for the $\psi^{+}_{\text{o}}$ at the notch tip vs. $\lambda$ curve for an intermediate loading speed of 0.062 mm/s. The $\psi^{+}_{\text{o}}$ evolution at the notch tip represents crack nucleation (point A), propagation (point A to point B) when $\psi^{+}_{\text{o}} > \psi^{+}_{cr}$, and subsequent end of propagation (point B) when $\psi^{+}_{\text{o}} < \psi^{+}_{cr}$. This agrees with the experimental observation in \citep{Shenmodeling}, which showed that at an intermediate loading speed, the crack nucleated and propagated and later arrested.}} 
    \label{EPS25validation}
\end{figure}

To test the model predictions for a different dynamically crosslinked soft polymer and a loading mode other than tension, we utilize the data reported by Shen and co-workers \citep{Shenmodeling} for the fracture of EPS25 vitrimer. Pure shear tests over a range of loading speeds were performed on EPS25 vitrimer specimens with sharp single-edge notches in \citep{Shenmodeling}. The specimen geometry is shown in Figure \ref{EPS25validation}A. The finite element mesh used for our simulations is shown in Figure \ref{EPS25validation}B. The crack extension vs. stretch ratio ($\lambda$) experimental data for a fast and a slow loading speed reported in \citep{Shenmodeling} is shown in Figure \ref{EPS25validation}C. A transition from high extensibility, i.e., no crack propagation (0.0225 mm/s) to fracture (0.09 mm/s) with increasing loading rates was observed as seen in Figure \ref{EPS25validation}C. We used the neo-Hookean function for mechanism $B$ with our same dynamic crosslink formulation. $G$ is set equal to 0 MPa in our model, and the deformation-associated material parameters from \citep{Shenmodeling} are used. In the experimental work, the authors report creating a sharp notch in the specimens using a blade. As the exact radius of the notch is not specified in \citep{Shenmodeling}, we assumed a very sharp notch radius of 100 $\mu$m. 

With the aim of modeling the initiation of crack propagation, only a VUMAT was used. A total of 28,716 C3D8 (three-dimensional, full integration) stress elements were used in the finite element mesh for the problem, as shown in Figure \ref{EPS25validation}B. $\psi^{+}_{cr}=\psi^{(B)}_{cr}$, $\mathit{W}_{cr}$ were evaluated as the values of $\psi^{+}_{\text{o}} = \psi^{(B)}_{\text{o}}$, $\mathit{W}$ at the notch tip at the time instant corresponding to the initiation of crack propagation for the 0.09 mm/s loading speed. Model predictions for $\psi^{+}_{\text{o}}$ at the notch tip vs. $\lambda$ curves for the fast and slow loading speeds are shown in Figure \ref{EPS25validation}D. The model predicts no crack propagation for the relatively slow 0.0225 mm/s loading speed due to $\psi^{+}_{\text{o}}$ at the notch tip not being able to exceed $\psi^{+}_{cr}$ as seen in Figure \ref{EPS25validation}D. This agrees with the experimental observation from \citep{Shenmodeling} in Figure \ref{EPS25validation}C. We also obtain the model prediction for the $\psi^{+}_{\text{o}}$ at the notch tip vs. $\lambda$ curve at an intermediate loading speed (0.062 mm/s). For this intermediate speed, $\psi^{+}_{\text{o}}$ at the notch tip exceeds $\psi^{+}_{cr}$ after a certain stretch (point A in Figure \ref{EPS25validation}E). However, due to subsequent significant network reconfiguration, the model predicts that $\psi^{+}_{\text{o}}$ drops below $\psi^{+}_{cr}$ after point B in Figure \ref{EPS25validation}E. The $\psi^{+}_{\text{o}}$ evolution at the notch tip represents crack nucleation (point A), propagation (point A to point B) when $\psi^{+}_{\text{o}} > \psi^{+}_{cr}$, and subsequent end of propagation (point B) when $\psi^{+}_{\text{o}} < \psi^{+}_{cr}$. The model prediction is in qualitative agreement with the experimental observation in \citep{Shenmodeling}, which showed that at an intermediate loading speed, the crack nucleated and propagated and later stopped propagating (arrested).

\subsection{SBR and NR: Failure surface predictions}

\begin{figure}[h!]
    \begin{center}
		\includegraphics[width=0.87\textwidth]{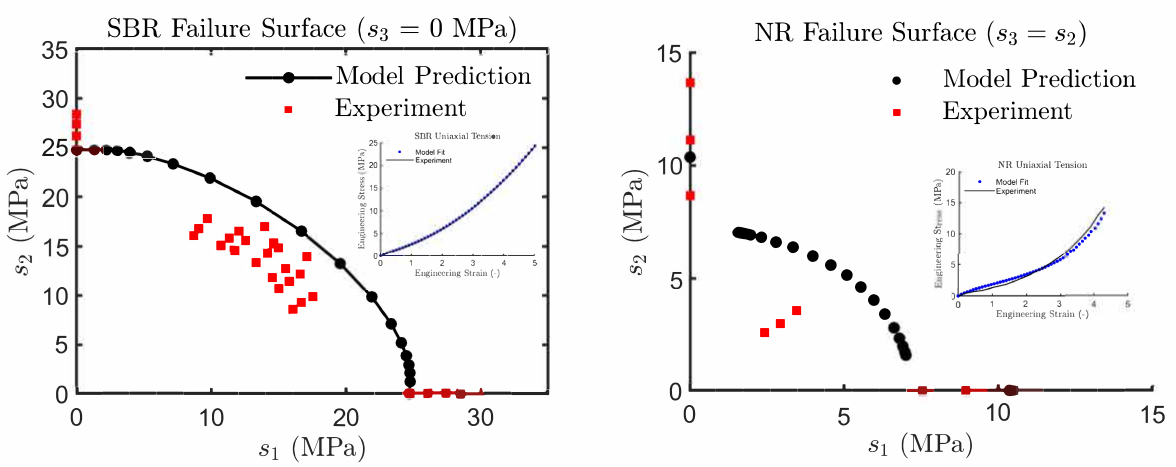}
	\end{center}
    \caption{{\textbf{Failure surface model predictions for SBR and NR.} The model predicts the failure surfaces for SBR ($s_3$ = 0 MPa) and NR ($s_3$ = $s_2$) reasonably well when its predictions are compared with the experimental surfaces from \citep{HAMDI2006994} and \citep{NRfailuresurface}. $s_{i}$ are the nominal principal stresses. Insets (deformation response): Incompressible Ogden (N=2) hyperelastic model and Gent model fit to the experimental uniaxial stress-strain curves from \citep{HAMDI2006994} for SBR and NR, respectively. Unique values of $\psi_{cr}$ were chosen as model parameters for both SBR and NR.}} 
    \label{SBRvalidation}
\end{figure}

Variational phase field models, which predict crack nucleation through minimization of a non-linear energy function without explicit criteria have been shown to be unable to capture the experimentally observed failure surfaces of various materials as discussed in the review article by Lopez and co-workers \citep{LOPEZPAMIES2025117520}. We compare failure surface predictions from our model, which is based on the method of virtual power and considers an independent criterion for damage initiation, with the experimental measurements for SBR from \citep{HAMDI2006994} and NR\footnote{A variational phase field model was shown in \citep{LOPEZPAMIES2025117520} to not be able to correctly predict the experimental data for SBR in \citep{HAMDI2006994} and NR in \citep{NRfailuresurface}. It showed unphysical predictions, specifically the multiaxial (biaxial for SBR, triaxial for NR) failure stress magnitudes to be larger than the uniaxial failure stress for both SBR and NR, with very significant deviations for NR.} from \citep{NRfailuresurface}. Specifically, our loading mode insensitive damage initiation criterion in the form of a unique critical free energy density $\psi^{+}_{cr}$ is used and validated. Due to the fast damage evolution of soft polymers, the failure occurs when $\psi^{+}=\psi^{+}_{cr}$. The deformation responses of SBR and NR were modeled using the incompressible Ogden (N=2) hyperelastic function and Gent hyperelastic function, respectively, fit to the experimental data from \citep{HAMDI2006994}. Unique $\psi^{+}_{cr}$ values were chosen based on the experimental failure data for both SBR and NR. Model predictions for the failure surfaces of SBR and NR using the proposed damage initiation criterion show good agreement with the experimental surfaces\footnote{It is prudent to note that the experimental fracture measurements in elastomers typically have a large variability and uncertainty, but the measures are useful to evaluate trends and obtain failure stress approximations within an order of magnitude.} as shown in Figure \ref{SBRvalidation}. 

\section{Closing remarks}

Soft polymers are found in nature and as engineering materials, including elastomers, viscoelastic polymers, polymer gels, and biological tissues. Their properties vary from rate-independent to rate-dependent. Current fracture toughness measures are non-unique for rate-dependent soft polymers for varying loading profiles and specimen geometries. We propose and show that a critical value of stress work $\mathit{W}_{cr}$ can quantify the fracture resistance of a certain class of soft polymers. $\mathit{W}_{cr}$ can be interpreted as the total energy per unit referential volume associated with the complete failure of an arbitrary material point under a loading mode subjected to sufficiently fast loading rates for damage initiation. 

We have developed a multi-mechanism, gradient-damage framework-based model to predict damage initiation, growth, and complete failure in soft polymers. The method of virtual power is used to formulate the gradient-damage framework. The energetic contribution to the critical stress work, $\psi^{+}_{cr}$, is proposed as a damage initiation criterion that accounts for material strength. The proposed damage model shows successful predictions of Polyborosiloxane's fracture in several independent experiments involving different specimen geometries and loading conditions. It is also able to predict the fracture of three elastomers, EPDM, SBR, NR, and another viscous soft polymer, EPS25. A transition in fracture profile from trumpet to single parabola-like was observed in the model predictions as the polymer viscous effects were removed. This is consistent with experimental observations, demonstrating the model's ability to predict the significantly varying physical phenomena observed during the fracture of soft polymers. The continuum scale model is developed based on microscopic physical insights. We show that macroscopic model parameters can be used to obtain relevant physical quantities at the microstructural process scales. From the macroscopic parameter $\psi^{+}_{cr}$, we estimated the value of subchain dissociation energy, which was verified by a set of independent experiments. In the future, more experiments with complex specimen geometries, loading rates, and loading modes can be performed to further study the approximate independence of the critical stress work from these variables. The field of damage and failure of soft materials is challenging and continuously evolving. Accurate and useful damage and fracture models that apply to a broad range of materials require good constitutive models that capture all important physics of damage initiation and growth, along with robust computational implementations.

\section*{Author Contributions}
\textbf{Aditya Konale}: Investigation, Methodology, Formal analysis, Data curation, Software, Visualization, Writing – original draft preparation, review \& editing.  \textbf{Vikas Srivastava}: Conceptualization, Investigation, Methodology, Supervision, Funding acquisition, Writing – original draft preparation, review \& editing.

\section*{Acknowledgments}
The authors would like to acknowledge the support from the Office of Naval Research (ONR), USA to V. Srivastava under grant numbers N00014-21-1-2815 and N00014-21-1-2670. The authors also thank Dr. James LeBlanc from the Naval Underwater Warfare Center for supporting the research.

\section*{Declaration of Interest}
The authors declare no competing interests.

\bibliography{SPF_arxiv_v4}

\newpage

\end{document}